\documentclass[aps,prb,twocolumn,preprintnumbers,showpacs,amsmath,floatfix,amssymb]{revtex4}

\usepackage{amsmath, graphicx, subfigure, bm, url, rotating, color, epsfig, dcolumn, multirow, longtable}

\providecommand{\tabularnewline}{\\}

\begin{document}

\title{Photoelectron properties of DNA and RNA bases from many-body
  perturbation theory}

\author{Xiaofeng Qian$^{1}$, Paolo Umari$^{2}$, and Nicola
  Marzari$^{1,3}$}

\affiliation{$^1$Department of Materials Science and Engineering,
  Massachusetts Institute of Technology, Cambridge, Massachusetts
  02139, USA}

\affiliation{$^2$Theory at Elettra Group, CNR-IOM Democritos,
  Basovizza-Trieste, Italy}

\affiliation{$^3$Department of Materials, University of Oxford, Oxford OX1 3PH, UK}

\date{\today}

\begin{abstract}
  
  The photoelectron properties of DNA and RNA bases are studied using
  many-body perturbation theory within the $GW$ approximation,
  together with a recently developed Lanczos-chain approach.
  Calculated vertical ionization potentials, electron affinities, and
  total density of states are in good agreement with experimental
  values and photoemission spectra. The convergence benchmark
  demonstrates the importance of using an optimal polarizability basis
  in the $GW$ calculations. A detailed analysis of the role of
  exchange and correlation in both many-body and density-functional
  theory calculations shows that while self-energy corrections are
  strongly orbital-dependent, they nevertheless remain almost constant
  for states that share the same bonding character. Finally, we report
  on the inverse lifetimes of DNA and RNA bases, that are found to
  depend linearly on quasi-particle energies for all deep valence
  states. In general, our $G_0W_0$-Lanczos approach provides an
  efficient yet accurate and fully converged description of
  quasiparticle properties of five DNA and RNA bases.
  
\end{abstract}


\maketitle 

\section{INTRODUCTION}

Understanding the photoelectron properties of DNA and RNA bases and
strands is of central importance to the study of DNA damage following
exposure to ultraviolet light or ionizing radiation~\cite{ColsonS95},
and to the development of fast DNA sequencing
techniques~\cite{ZwolakD05} and DNA and RNA-based molecular
electronics and sensors~\cite{PorathBdD00,KawaiKOM09}. Extensive
experimental efforts~\cite{HushC75, DoughertyYVAM78, ChoiLK05,
  TrofimovSKPHK06, SchwellJBL08, ZaytsevaTSPFRCP09, KostkoBKA10} have
been made since the 1970's to measure the photoelectron properties of
DNA and RNA bases. Meanwhile, theoretical calculations on their
ionization potentials and electron affinities have been carried out
using density-functional theory (DFT) and high-level quantum chemistry
methods~\cite{RussoTG00, TrofimovSKPHK06, Roca-SanjuanRMS06,
  Roca-SanjuanMSR08, ZaytsevaTSPFRCP09, BravayaKDLAK10}. However, the
results from DFT calculations are highly dependent on
exchange-correlation functionals, and quantum chemistry methods,
though more accurate, require considerably more computational
effort. In contrast, many-body perturbation theory within Hedin's $GW$
approximation~\cite{Hedin65, HedinL69} presents a unique framework
that allows access to both quasi-particle (QP) energies and lifetimes
on the same footing.  This method has been successfully applied to
quasi-one-dimensional(1D), two-dimensional (2D) and three-dimensional
(3D) semiconductors, insulators, and metals~\cite{HybertsenL86,
  RojasGN95, RiegerSWRG99, CampilloPRZE99, SpataruCRBEL01, OnidaRR02},
and very recently to molecular systems~\cite{DoriMKSKU06, TiagoKHR08,
  PalummoHSBR09, UmariSB09, UmariSB10, StenuitCPFPGU10, UmariQMSGB11,
  RostgaardJT10, BlaseAO11, FaberAORB11}.

In this work, we present the {\em entire} QP spectrum of DNA and RNA
bases using many-body perturbation theory within Hedin's $GW$
approximation, obtained with a recently developed approach that is
particularly effective in reaching numerical
convergence~\cite{UmariSB09,UmariSB10}. In the $GW$ approximation, the
self-energy operator is expressed as a convolution of the QP Green's
function $G$ with the screened Coulomb interaction $W$. Therefore, at
increasing system sizes (as is the case for the present work) two
computational challenges arise: (i) first, a large basis set has to be
adopted to represent operators such as polarizability, and (ii) the
calculation of the irreducible dynamical polarizability and that of
the self-energy require sums over single-particle conduction states
that converge very slowly.  We overcome these two obstacles through
(i) the use of optimal basis sets for representing the polarization
operators~\cite{UmariSB09} and (ii) the use of a Lanczos-chain
algorithm~\cite{UmariSB10} to avoid explicit sums over empty
single-particle states. In addition, the $G_0W_0$ approximation is
adopted, in which the dynamical polarizability is calculated within
the random-phase approximation and the QP Green's function is replaced
by its unperturbed single-particle counterpart.  This approach is
implemented in the open-source {\sc Quantum-ESPRESSO}
distribution~\cite{ESPRESSO10}. It is applied here to achieve fully
converged QP spectra and inverse lifetimes of the five isolated DNA
and RNA bases and to investigate the important but distinct roles of
exchange and correlation in the $G_0W_0$ self-energy corrections.

Photoelectron properties of DNA and RNA bases using many-body $GW$
have not been reported until a very recent study by Faber {\it et
  al.}\cite{FaberAORB11}. The work by Faber {\it et al.} presents a
many-body $GW$ study on QP energies (including ionization potentials
and electron affinities) of DNA and RNA bases at several levels of
self-consistency within the $GW$ approximation. Their calculations
were based on the conventional implementation of the $GW$ method using
a localized basis set and a direct sum-over-states approach, and it
demonstrated that self-consistent GW calculations indeed further
improve the results of $G_0W_0$ (one-shot $GW$) calculations. Although
the localized basis set can significantly improve the computational
efficiency and a direct sum-over-states approach can be easily
implemented, both of them could have several potential drawbacks, and
can introduce large errors in QP energies. One may solve the former
issue by systematically increasing the size of basis sets; however,
there is no simple solution to the convergence problem introduced by
the direct sum-over-state approach. Second, dipole-bound conduction
states will not be obtained from localized basis sets due to their
highly diffuse character in the vacuum region. In fact, only electron
affinities of covalent-bound conduction states were reported in the
Faber work. Therefore, it would be desirable to calculate $GW$ QP
energies in a plane-wave basis without suffering from the above
issues, which is one of the subjects of this research.

The paper is organized as follows.  Computational details of our
calculations are given in Sec.~\ref{SEC:COMP}. Real-space
representations of optimal polarizability basis are displayed in
Sec.~\ref{SEC:BASIS}. We then report convergence benchmark in
Sec.~\ref{SEC:CONVERGENCE}. In Sec.~\ref{SEC:IPEA}, we present QP
energies and inverse lifetimes as well as the entire QP spectra for all five
DNA and RNA bases, including guanine (G), adenine (A), cytosine (C),
thymine (T), and uracil (U). Vertical ionization potentials (VIPs) and
vertical electron affinities (VEAs) are compared to experimental
data and other theoretical results. Two types of VEAs are reported
using plane-wave basis, including valence-bound (VB, also called
covalent-bound) VEA and dipole-bound (DB) VEAs.  In
Sec.~\ref{SEC:ROLE}, we reveal the role of exchange and correlation in
$GW$ self-energy corrections to the DFT Kohn-Sham eigenvalues.
Finally, we summarize our work in Sec.~\ref{SEC:SUMMARY}.

\section{Computational details }
\label{SEC:COMP}

\begin{figure}[thpb]\centering
  \includegraphics[width=0.95\columnwidth]{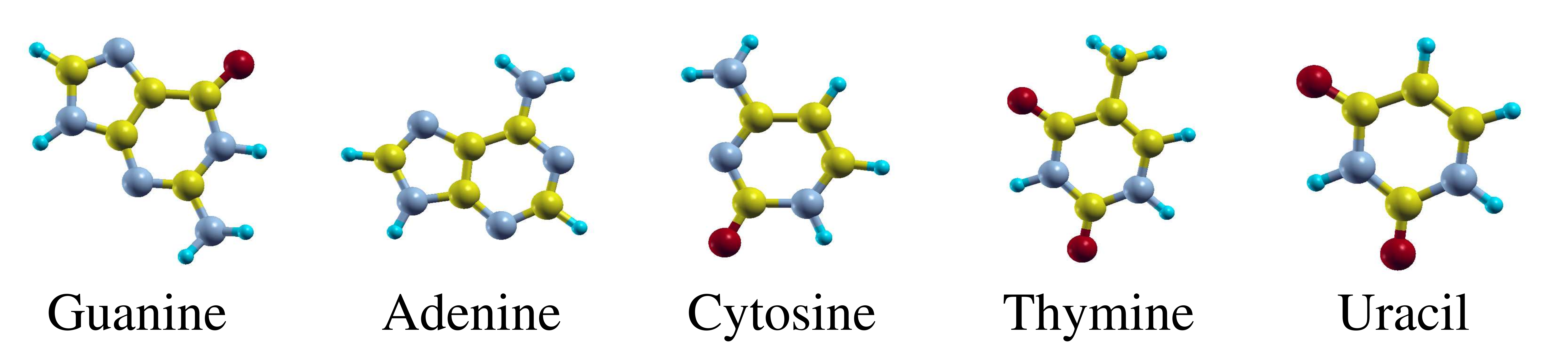}
  \caption{(color online). Ground-state structures of five DNA and RNA
    bases including G9K-guanine, adenine, C1-cytosine, thymine, and
    uracil.}
  \label{FIGURE:STRUCTURE}
\end{figure}

Ground-state DFT calculations are performed in a cubic supercell of
$18.0^3$~\AA$^3$, using the Perdew-Burke-Ernzerhof's (PBE)
exchange-correlation functional, Troullier-Martins's norm-conserving
pseudopotentials, and a plane-wave basis set with a cutoff of 544 eV.
Structures are optimized with a residual force threshold of 0.026
eV/\AA. A truncated Coulomb potential with radius cutoff of 7.4 {\AA}
is employed to remove artificial interactions from periodic
images. The vacuum level is corrected by an exponential fitting of
$E_{\rm HOMO}$ with respect to the supercell volume. The
polarizability basis sets have been obtained using a parameter $E^*$
of 136.1 eV and a threshold $q^*$ of 0.1 a.u., giving an accuracy of
0.05 eV for the calculated QP energies ($E^*$ and $q^*$ will be
explained in the next section). The final accuracy including the
errors from the analytic continuation is about 0.05 to 0.1 eV. The
structures of five DNA and RNA bases are shown in
Fig.~\ref{FIGURE:STRUCTURE}. Here the effect of gas-phase tautomeric
forms~\cite{BravayaKDLAK10} of guanine and cytosine on QP properties
are beyond the scope of this work, and we only focus on the G9K form
of guanine and the C1 form of cytosine~\cite{BravayaKDLAK10}.

\section{Optimal polarizability basis}
\label{SEC:BASIS}

The key quantity in many-body $GW$ calculations is the irreducible
dynamic polarizability $\hat{P}_0$ in the random-phase approximation:
\begin{equation}
  \hat{P}_0(\omega) =
  -i  
  \sum_{v,c} \frac{|\psi_v\psi_c\rangle \langle\psi_c\psi_v|}
  {\omega + i\eta - (\varepsilon_{c} - \varepsilon_{v})}, 
  \label{P0}
\end{equation}
where $\eta$ is an infinitesimal positive real number.
$|\psi_v\psi_c\rangle$ denotes the direct product of a valence state
$\psi_v$ and a conduction state $\psi_c$ in real space and $\psi_v$
and $\psi_c$ are considered to be real.  A strategy was proposed in
Refs.~\onlinecite{UmariSB09} and \onlinecite{UmariSB10} for obtaining
a compact basis set, referred to as {\it optimal polarizability
  basis}, to represent $\hat{P}_0$ at all frequencies. First, we
consider the frequency average of $\hat{P}_0(\omega)$ which
corresponds to the element at time $t=0$ of its Fourier transform
$\tilde{\hat{P}}_0(t)$, without considering the constant ($-i$):
\begin{equation}
  \tilde{\hat{P}}_0(t=0) =
  \sum_{v,c} |\psi_v\psi_c\rangle \langle\psi_c\psi_v|.
  \label{P1}
\end{equation}
We note that $\tilde{\hat{P}}_0(t=0)$ is positive definite. Then, the
{\it optimal polarizability basis}, $\{\Phi_{\mu}\}$, is built from
the most important eigenvectors of $\tilde{\hat{P}}_0(t=0)$,
corresponding to the largest eigenvalues $q_{\mu}$ above a given
threshold $q^*$:
\begin{equation}
  \tilde{\hat{P}}_0(t=0) \; | \Phi_{\mu}
  \rangle = q_{\mu} | \Phi_{\mu} \rangle.
  \label{Peigen}
\end{equation}
It must be noted that this does not require any explicit calculation
of empty (i.e., conduction) states as we can use the closure relation:
\begin{equation}
{\hat P}_c=1-{\hat P}_v,
\label{closure}
\end{equation}
together with an iterative diagonalization scheme.  However, the
latter procedure would build polarizability basis sets which are larger
than what is necessary for a good convergence of the quasi-particle
energy levels.  This stems from treating all the one-particle
excitations on the same footing, independent of their energy.  A
practical solution would be to limit the sum in Eq.~(\ref{P1}) on the
conduction states below a given energy cutoff $E^*$:
\begin{equation}
  \tilde{\hat{P}}'_0 =
  \sum_{v,c}^{\epsilon_c<E^*} |\psi_v\psi_c\rangle \langle\psi_c\psi_v|.
  \label{P2}
\end{equation}
However, limiting the sum over the empty states laying in the lower
part of the conduction manifold does not allow to use the closure
relation alluded to above.

Thus, to keep avoiding the calculation of empty states we replace them
in Eq.~(\ref{P2}) with a set of plane waves $\{ {\mathbf G} \}$ with
their kinetic energies lower than $E^*$, which are first projected
onto the conduction manifold using Eq.~(\ref{closure}) and then
orthonormalized. We indicate these {\it augmented plane-waves} as $\{
\tilde{\mathbf G} \}$ and arrive at the following modified operator:
\begin{equation}
  \tilde{\hat{P}}''_0 =
  \sum_{v,\tilde{\mathbf G}} |\psi_v\tilde{\mathbf G}\rangle \langle\tilde{\mathbf G}\psi_v|,
  \label{P3}
\end{equation}
which is also positive-definite.  An optimal polarizability basis
$\{\Phi_{\mu}\}$ is finally obtained by replacing
$\tilde{\hat{P}}_0(t=0)$ in Eq.~(\ref{Peigen}) with
$\tilde{\hat{P}}''_0$.

It should be stressed that the above approximation is used only for
obtaining a set of optimal basis vectors for representing the
polarization operators and not for the actual calculation of the
irreducible dynamic polarizability at finite frequency in
Eq.~(\ref{P0}); the latter is performed using a Lanczos-chain
algorithm \cite{UmariSB10}.  Moreover, due to the completeness of the
eigenvectors of $\tilde{\hat{P}}''$, for any value of $E^*$ the $GW$
results will converge to the same values by lowering the threshold
$q^*$, and eventually reach the same results as those obtained by
directly using a dense basis of plane-waves. However, compared to the
pure plane-waves which are completely delocalized in real space, the
optimal polarizability basis is particularly convenient for isolated
systems since the most important eigenvectors of $\tilde{\hat{P}}''_0$
will be mostly localized in the regions with higher electron density.
Thus, converged results can be obtained using much smaller
optimal-polarizability basis sets than plane-waves basis sets.

\begin{figure}[thpb]\centering
  \includegraphics[width=2.8in]{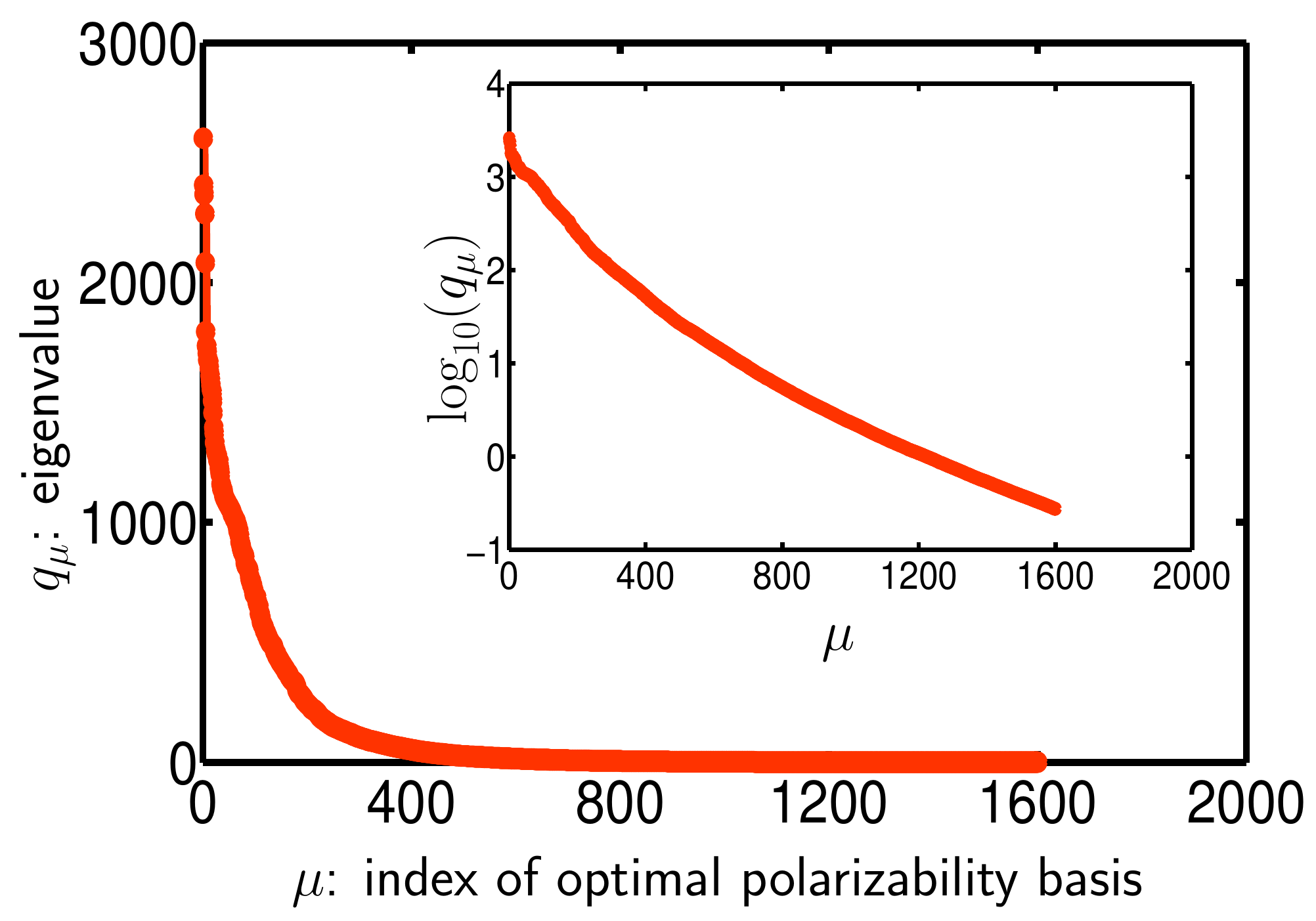}
  \caption{(Color online) Eigenvalue distribution of the optimal
    polarizability basis for cytosine. The inset plot shows the
    eigenvalues in a log scale.}
  \label{FIG:POLARBASIS_EIGENVALUE}
\end{figure}

\begin{figure}[thpb]\centering
  \includegraphics[width=0.98\columnwidth]{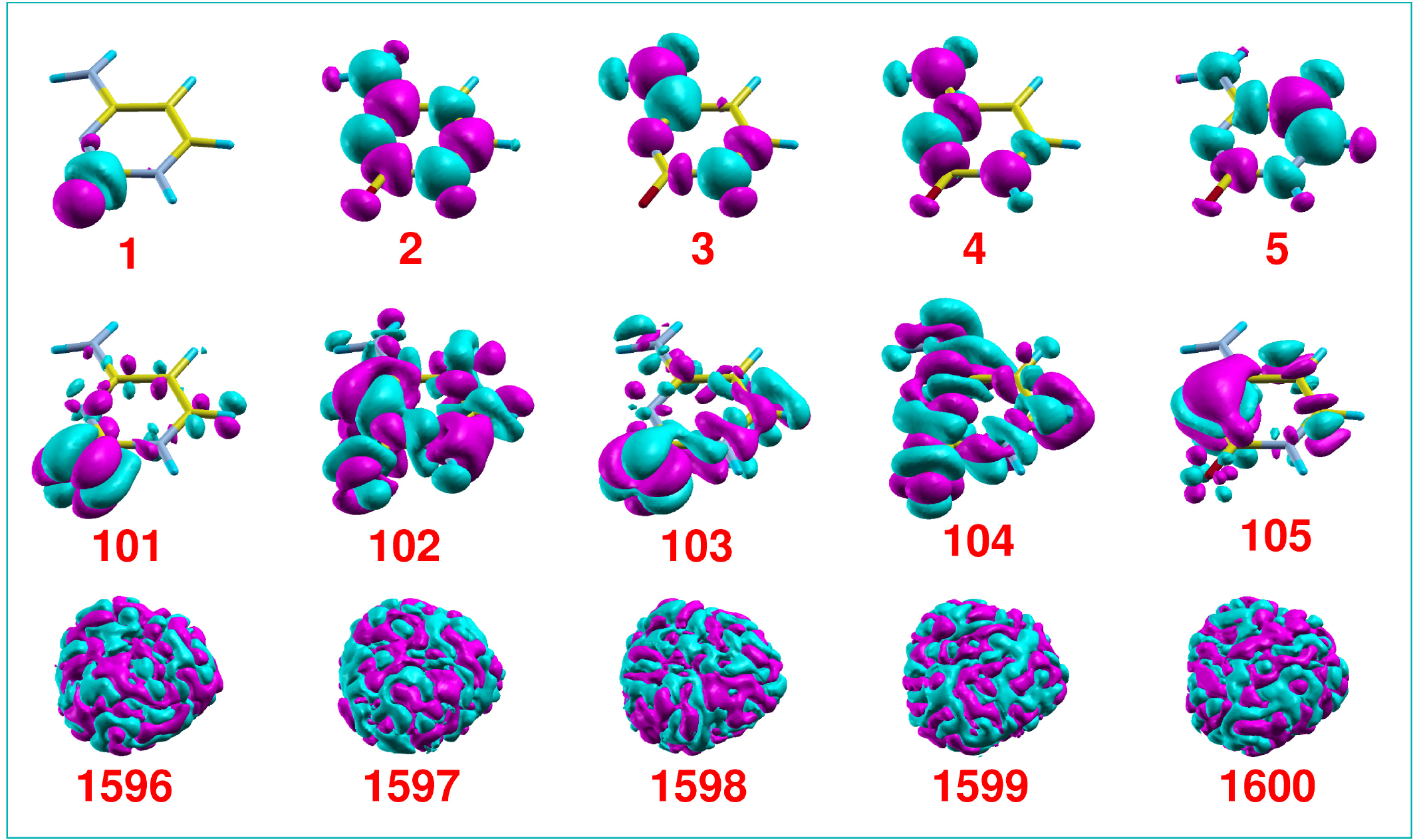}
  \caption{(color online). Real-space representation of
    optimal-polarizability basis elements for cytosine, labeled with
    their eigenvalue indexes. Due to the delocalized nature of the
    optimal basis in the third row, the images in the third row were
    generated with a smaller isovalue and shown at a larger scale than
    those in the first two rows.}
  \label{FIG:POLARBASIS_ORBITAL}
\end{figure}

Now, we want to have a closer look at the optimal polarizability
basis. The eigenvalue distribution of $\tilde{\hat{P}}''_0$ for
cytosine is displayed in Fig.~\ref{FIG:POLARBASIS_EIGENVALUE}. We only
show the largest 1600 eigenvalues with $E^*=136.1$ eV in the plot,
since these provide well converged results. It is clearly seen that
the eigenvalues of the optimal polarizability basis decay
exponentially and change by almost four orders of magnitude from the
first to the last basis. In Fig.~\ref{FIG:POLARBASIS_ORBITAL} we show
the real-space representations of a few selected elements. The first
five, corresponding to the five largest eigenvalues, are strongly
localized around the chemical bonds of the molecule.  The second row
contains five elements which are more delocalized, and those in the
last row are completely delocalized. This indicates that even though
localized optimal bases like those shown in the first two rows can be
easily captured by localized basis-sets, the delocalized ones with
smaller eigenvalues $q_{\mu}$ (like those in the last row) are more
difficult to capture if diffuse functions are not employed.

\section{CONVERGENCE BENCHMARK}
\label{SEC:CONVERGENCE}

\begin{figure}[thpb]\centering
  \includegraphics[width=0.95\columnwidth]{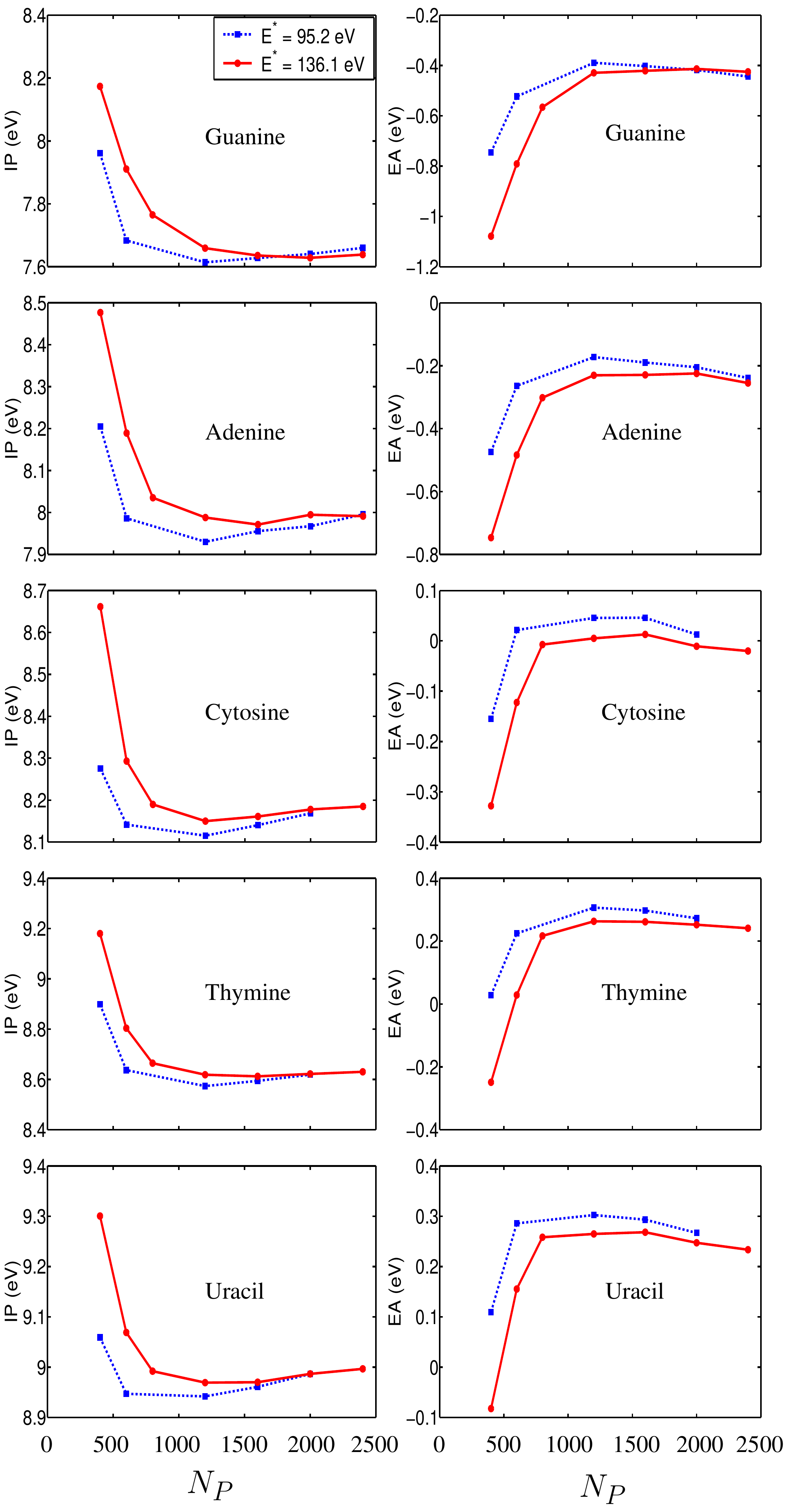}
  \caption{(color online). Convergence benchmark of VIP and VEA of
    five DNA and RNA bases with respect to the number of
    optimal-polarizability basis elements $N_P$, and augmented
    plane-wave cutoff $E^*$. Results using $E^*=$ 95.2 and 136.1 eV
    are plotted in dashed-blue lines and solid-red lines,
    respectively.}
  \label{FIG:CONVERGENCE}
\end{figure}

The number of optimal-polarizability basis elements $N_P$ and the
energy cutoff of the augmented plane-waves $E^*$ are two critical
parameters used in our $G_0W_0$ calculations to achieve both
efficiency and accuracy. Therefore, we performed a series of
calculations to benchmark the convergence with respect to these two
parameters. In Fig.~\ref{FIG:CONVERGENCE}, we present the convergence
behavior of VIPs and VB-VEAs of five DNA and RNA bases for the
highest-occupied molecular orbital (HOMO) and the lowest-unoccupied
molecular orbital (LUMO), respectively: ${\rm VIP} \equiv - {\rm
  Re}(\varepsilon_{\rm HOMO}^{\rm QP})$ and ${\rm VEA} \equiv - {\rm
  Re}(\varepsilon_{\rm LUMO}^{\rm QP})$.  We find that for both VIPs
and VEAs convergence within $0.1$ eV is achieved with $\sim600$
optimal basis elements for $E^*= 95.2$ eV and with $\sim750$ optimal
basis elements for $E^*= 136.1$ eV.  Indeed, similar trends were
reported in Ref.~\onlinecite{UmariSB09}. VIPs and VEAs reported in the
following sections are calculated using the most strict parameters
($N_P=2400$ and $E^*=136.1$ eV).

The above benchmark indicates that, if basis-sets and conduction
states in DFT calculations are not properly tested, one could easily
obtain non-converged results from $G_0W_0$ calculations, resulting in
higher VIPs and lower VEAs for all five bases. We also note that the
choice of $N_P$ and $E^*$ remains the same for all the DNA and RNA
bases, indicating portability for these parameters.

\section{IONIZATION POTENTIALS AND ELECTRON AFFINITIES}
\label{SEC:IPEA}

\begin{figure}[thpb]\centering
  \includegraphics[width=0.8\columnwidth]{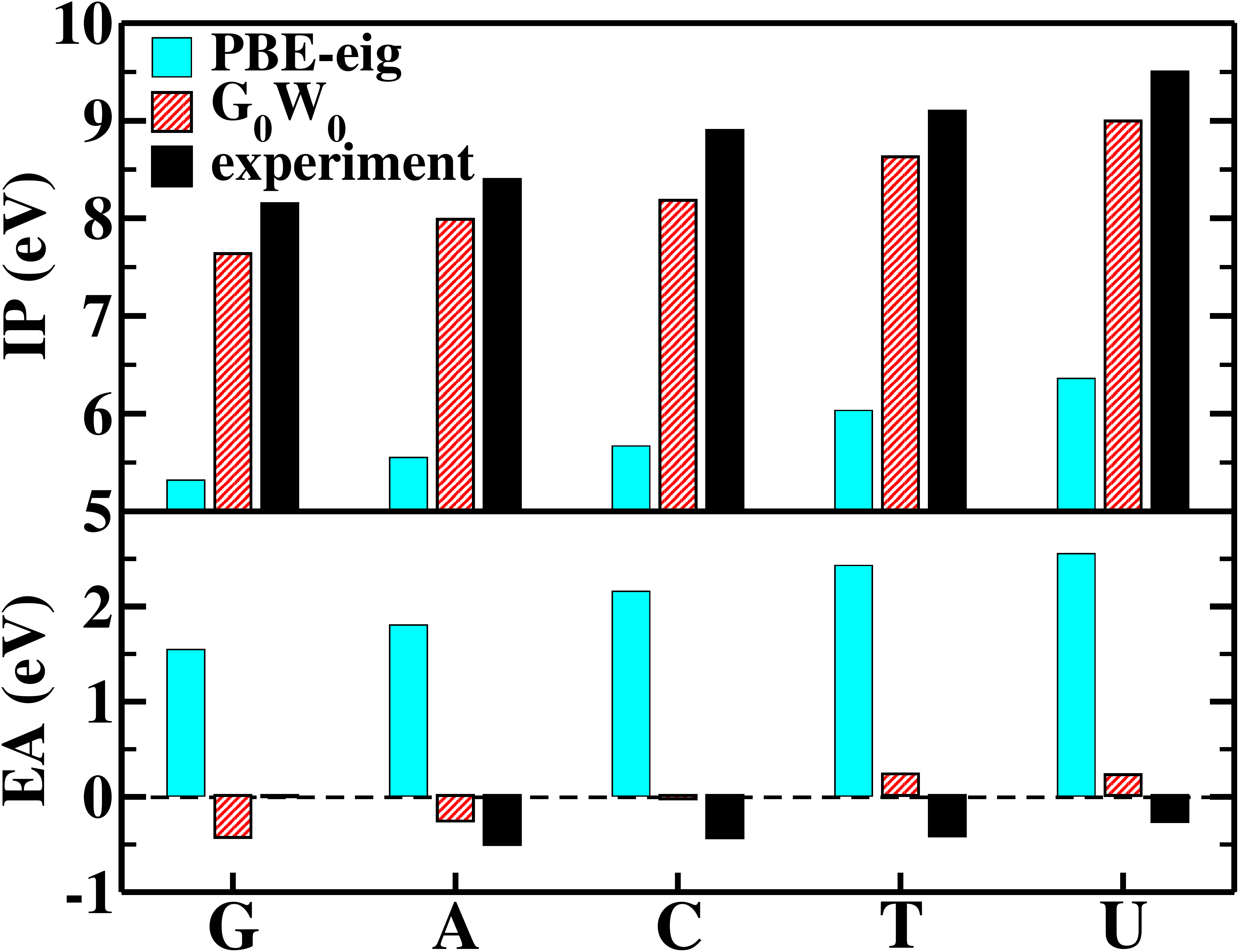}
  \caption{(color online). VIP and VB-VEA of five DNA and RNA bases
    from our DFT and $G_0W_0$ calculations. Here we adopt the mean
    values of various experimental data listed in
    Table~\ref{TABLE1}. The experimental ranges are $8.0\sim8.3$,
    $8.3\sim8.5$, $8.8\sim8.9$, $9.0\sim9.2$, and $9.4\sim9.6$ eV for
    G, A, C, T, and U, respectively.}
  \label{FIGURE:VIP_VEA}
\end{figure}

VIPs and VB-VEAs from our $G_0W_0$ calculations and experimental data
are shown in Fig.~\ref{FIGURE:VIP_VEA} for all five bases, together
with the DFT-PBE eigenvalues for the HOMO and LUMO levels. Only the
mean values of experimental VIPs and VEAs are plotted in
Fig.~\ref{FIGURE:VIP_VEA}. $G_0W_0$ dramatically improves VIPs and
VEAs compared to DFT-PBE eigenvalues, providing VIPs of 7.64, 7.99,
8.18, 8.63 and 8.99 eV and VEAs of $-0.43$, $-0.25$, $-0.02$, $0.24$,
and $0.23$ eV for G, A, C, T, and U, respectively. The experimental
VIPs are compiled in Table~\ref{TABLE1}, and span a range of
$8.0\sim8.3$, $8.3\sim8.5$, $8.8\sim8.9$, $9.0\sim9.2$, and
$9.4\sim9.6$ eV for G, A, C, T, and U. Compared to the mean values of
experimental VIPs, the mean absolute error of the calculated VIPs for
all five bases is 0.52 eV.  Furthermore, experimental VB-VEAs are
negative for all five bases, indicating that excited $\pi^*$ states
are unstable upon electron attachment. This leads to challenging
measurements of VEAs and a wide range of measured
values~\cite{RussoTG00, Roca-SanjuanMSR08} listed in
Table~\ref{TABLE1}: $-0.56\sim-0.45$, $-0.55\sim-0.32$,
$-0.53\sim-0.29$, and $-0.30\sim-0.22$, for A, C, T, and U. Compared
to the mean values of experimental VEAs, the mean absolute errors of
the calculated VEAs for four bases is 0.45 eV. Interestingly, the VEA
of guanine has never been measured successfully, possibly due to a
large negative value. This is clearly reflected in our calculated
$G_0W_0$ VEA of $-0.43$ eV, which is the most negative one among all
five bases. Even though the $G_0W_0$ VEAs of thymine and uracil are
slightly positive, the trend for all the calculated VEAs agrees well
with experiments. In addition, the DFT-PBE HOMO-LUMO gaps for the five
bases are about 45\% of the $G_0W_0$ gaps. This is in agreement with
previous observations that DFT with the local density approximation
(LDA) or the generalized gradient approximation (GGA) of
exchange-correlation functionals usually underestimates by 30-50\% the
true QP energy gap~\cite{GodbySS86, GruningMR06}.

\begin{table*}
  \caption{Vertical ionization potentials and vertical electron affinities for several low-lying Kohn-Sham eigenstates close to the HOMO and LUMO levels obtained from negative DFT-PBE eigenvalues and $G_0W_0$(PBE) in comparison with other $GW$ and quantum chemistry calculations and experimental data. CASPT2: complete active space with second-order perturbation theory; CCSD(T): coupled-cluster with singles, doubles, and perturbative triple excitations; EOM: equation of motion ionization potential coupled-cluster. Only the valence-bound vertical electron affinities are shown in this table. The experimental mean values are taken as the reference in the calculations of mean absolute error(MAE) for both LUMO and HOMO levels.}
  \begin{ruledtabular}
    \begin{tabular}{crrrrrrr}
      \noalign{\smallskip}
 & DFT-PBE\footnotemark[1] & \text{$G_0W_0$(PBE)}\footnotemark[1] & $G_0W_0$(LDA)\footnotemark[2] 
 & $GW$(LDA)\footnotemark[2] & CASPT2\footnotemark[3]$^,$\footnotemark[4]/CCSD(T)\footnotemark[3]$^,$\footnotemark[4]
 & EOM\footnotemark[5] & Experiment\footnotemark[6]$^,$\footnotemark[7]$^,$\footnotemark[8]$^,$\footnotemark[9]$^,$\footnotemark[10] \\\noalign{\smallskip}
\hline
\noalign{\smallskip}
\multirow{8}{*}{G} & [LUMO] 1.12~($\pi$) & --0.43 & --1.04 & --1.58 & --1.14\footnotemark[3]/ &  & \tabularnewline \noalign{\smallskip}
 & [HOMO] 5.32~($\pi$) &  7.64 & 7.49 & 7.81 & 8.09\footnotemark[4]/8.09\footnotemark[4] & 8.15 & 8.0$\sim$8.3\footnotemark[6]/8.30\footnotemark[9]/8.26\footnotemark[10]\tabularnewline
 & 5.88~($n$)   &  8.67 & 8.78 & 9.82 &  9.56\footnotemark[4]/ &  9.86 & 9.90\footnotemark[9]/9.81\footnotemark[10]\tabularnewline
 & 6.37~($n$)   &  9.38 &      &      &  9.61\footnotemark[4]/ & 10.13 & \tabularnewline
 & 7.04~($\pi$) &  9.43 &      &      & 10.05\footnotemark[4]/ & 10.29 & \tabularnewline
 & 6.94~($\pi$) &  9.48 &      &      & 10.24\footnotemark[4]/ & 10.58 & 10.45~($n$)\footnotemark[9]/10.36\footnotemark[10]\tabularnewline
 & 7.76~($\pi$) & 10.37 &      &      & 10.90\footnotemark[4]/ & 11.38 & 11.15\footnotemark[9]/11.14\footnotemark[10]\tabularnewline
 & 7.64~($n$)   & 10.57 &      &      &                        &  & \tabularnewline
\noalign{\smallskip}
\hline 
\noalign{\smallskip}
\multirow{7}{*}{A} & [LUMO] 1.81~($\pi$) & --0.25 & --0.64 & --1.14 & --0.91\footnotemark[3]/ &  & --0.56$\sim$--0.45\footnotemark[7]\tabularnewline \noalign{\smallskip}
 & [HOMO] 5.55~($\pi$) &  7.99 & 7.90 & 8.22 &  8.37\footnotemark[4]/8.40\footnotemark[4] & 8.37 & 8.3$\sim$8.5\footnotemark[6]/8.47\footnotemark[8]\tabularnewline
 & 5.89~($n$)   &  8.80 & 8.75 & 9.47 &  9.05\footnotemark[4]/ &  9.37 &  9.45\footnotemark[8]\tabularnewline
 & 6.65~($\pi$) &  9.06 &      &      &  9.54\footnotemark[4]/ &  9.60 &  9.54\footnotemark[8]\tabularnewline
 & 6.74~($n$)   &  9.71 &      &      &  9.96\footnotemark[4]/ & 10.42 & 10.45\footnotemark[8]\tabularnewline
 & 7.22~($\pi$) &  9.78 &      &      & 10.38\footnotemark[4]/ & 10.58 & 10.51\footnotemark[8]\tabularnewline
 & 7.58~($n$)   & 10.65 &      &      & 11.06\footnotemark[4]/ & 11.47 & 11.35\footnotemark[8]\tabularnewline
\noalign{\smallskip}
\hline 
\noalign{\smallskip}
\multirow{7}{*}{C} & [LUMO] 2.16~($\pi$) & --0.02 & --0.45 & --0.91 & --0.69\footnotemark[3]/--0.79\footnotemark[3] &  & --0.55$\sim$--0.32\footnotemark[7]\tabularnewline \noalign{\smallskip}
 & [HOMO] 5.67~($\pi$)  &  8.18 & 8.21 &  8.73 & 8.73\footnotemark[4]/8.76\footnotemark[4] & 8.78 & 8.8$\sim$9.0\footnotemark[6]/8.89\footnotemark[8]\tabularnewline
 & 5.63~($n$)    &  8.50 & 8.80 &  9.89 &  9.42\footnotemark[4]/ &  9.65 &  9.45\footnotemark[9]/9.55\footnotemark[8]\tabularnewline
 & 6.28~($\pi$)  &  8.94 & 8.92 &  9.52 &  9.49\footnotemark[4]/ &  9.55 &  9.89\footnotemark[8]\tabularnewline
 & 6.38~($n$)    &  9.39 & 9.38 & 10.22 &  9.88\footnotemark[4]/ & 10.06 & 11.20\footnotemark[8]\tabularnewline
 & 8.44~($\pi$)  & 11.08 &      &       & 11.84\footnotemark[4]/ & 12.28 & 11.64\footnotemark[8]\tabularnewline
 & 9.27~($\pi$)  & 11.98 &      &       & 12.71\footnotemark[4]/ & 13.27 & 12.93~($\sigma$, $\pi$)\footnotemark[8]\tabularnewline
\noalign{\smallskip}
\hline 
\noalign{\smallskip}
\multirow{7}{*}{T} & [LUMO] 2.43~($\pi$) & 0.24& --0.14 & --0.67 & --0.60\footnotemark[3]/--0.65\footnotemark[3] &  & --0.53$\sim$--0.29\footnotemark[7]\tabularnewline \noalign{\smallskip}
 & [HOMO] 6.03~($\pi$)  & 8.63  & 8.64 &  9.05 &  9.07\footnotemark[4]/9.04\footnotemark[4] & 9.13 & 9.0$\sim$9.2\footnotemark[6]/9.19\footnotemark[8]\tabularnewline
 & 6.12~($n$)    & 8.94  & 9.34 & 10.41 &  9.81\footnotemark[4]/ & 10.13 &  9.95$\sim$10.05\footnotemark[6]/10.14\footnotemark[8]\tabularnewline
 & 6.80~($\pi$)  &  9.52 &      &       & 10.27\footnotemark[4]/ & 10.52 & 10.39$\sim$10.44\footnotemark[6]/10.45\footnotemark[8] \tabularnewline
 & 6.93~($n$)    &  9.77 &      &       & 10.49\footnotemark[4]/ & 11.04 & 10.80$\sim$10.88\footnotemark[6]/10.89\footnotemark[8] \tabularnewline
 & 8.79~($\pi$)  & 11.53 &      &       & 12.37\footnotemark[4]/ & 12.67 & 12.10$\sim$12.30\footnotemark[6]/12.27\footnotemark[8] \tabularnewline
\noalign{\smallskip}
\hline 
\noalign{\smallskip}
\multirow{7}{*}{U}  & [LUMO] 2.55~($\pi$) & 0.23 & --0.11 & --0.64 & --0.61\footnotemark[3]/--0.64\footnotemark[3] &  & --0.30$\sim$--0.22\footnotemark[7]\tabularnewline \noalign{\smallskip}
 & [HOMO] 6.36~($\pi$)  &  8.99 &  9.03 &  9.47 &  9.42\footnotemark[4]/9.43\footnotemark[4] & & 9.4$\sim$9.6\footnotemark[6]\tabularnewline
 & 6.14~($n$)    &  9.07 &  9.45 & 10.54 &  9.83\footnotemark[4]/ &  & 10.02$\sim$10.13\footnotemark[6]\tabularnewline
 & 7.00~($\pi$)  &  9.68 &  9.88 & 10.66 & 10.41\footnotemark[4]/ &  & 10.51$\sim$10.56\footnotemark[6]\tabularnewline
 & 6.92~($n$)    &  9.96 & 10.33 & 11.48 & 10.86\footnotemark[4]/ &  & 10.90$\sim$11.16\footnotemark[6]\tabularnewline
 & 9.17~($\pi$)  & 11.90 &       &       & 12.59\footnotemark[4]/ &  & 12.50$\sim$12.70\footnotemark[6]\tabularnewline
\noalign{\smallskip}
\hline 
\noalign{\smallskip}
\multirow{2}{*}{MAE} & [LUMO]~2.64~($\pi$) & 0.45 & 0.14 & 0.44 & 0.30/0.33 &  & \tabularnewline
 & [HOMO]~3.02~($\pi$) & 0.52 & 0.56 & 0.15 & 0.07/0.07 & 0.05 & \tabularnewline
\noalign{\smallskip}
\end{tabular}
\footnotetext[1]{~This work.}
\footnotetext[2]{~Ref.~\onlinecite{FaberAORB11}.}
\footnotetext[3]{~Ref.~\onlinecite{Roca-SanjuanMSR08}.}
\footnotetext[4]{~Ref.~\onlinecite{Roca-SanjuanRMS06}.}
\footnotetext[5]{~Ref.~\onlinecite{BravayaKDLAK10}.}
\footnotetext[6]{~Collected in Ref.~\onlinecite{Roca-SanjuanRMS06}.}
\footnotetext[7]{~Collected in Ref.~\onlinecite{Roca-SanjuanMSR08}.}
\footnotetext[8]{~Ref.~\onlinecite{TrofimovSKPHK06}.}
\footnotetext[9]{~Ref.~\onlinecite{DoughertyYVAM78}.}
\footnotetext[10]{~Ref.~\onlinecite{ZaytsevaTSPFRCP09}.}
\end{ruledtabular}
\label{TABLE1}
\end{table*}

\begin{figure}[thpb]\centering
  \includegraphics[width=0.65\columnwidth]{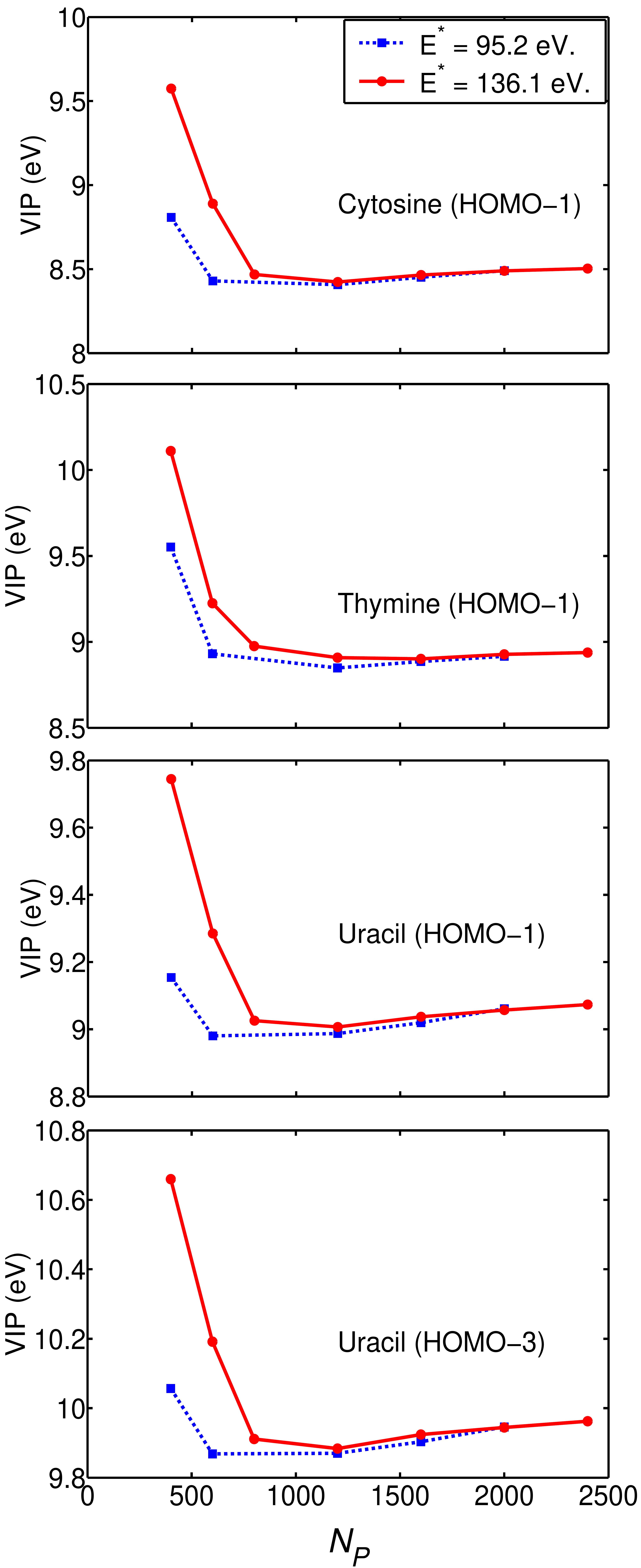}
  \caption{(color online). Convergence behavior of vertical ionization
    potentials of several $n$ states in DNA and RNA bases with respect
    to the number of optimal polarizability basis, $N_P$, and
    augmented plane-wave cutoff, $E^*$. These states are cytosine's
    HOMO-1 state, thymine's HOMO-1 state, and uracil's HOMO-1 and
    HOMO-3 states. Results using $E^*=$ 95.2 and 136.1 eV are plotted
    in dashed-blue lines and solid-red lines, respectively.}
  \label{FIG:CONVERGENCE2}
\end{figure}

We further compare several low-lying $G_0W_0$ VIPs and their
excitation characters with experimental and other theoretical results
and assignments. First, as shown in Table~\ref{TABLE1}, both $G_0W_0$
VIPs and their orbital assignments agree well with experiments and
other theoretical works for all the five bases, where the
corresponding excitation character is either $\pi$ or $n$ (lone
pair). Second, our $G_0W_0$ VIPs, especially those corresponding to
the five HOMO levels, are in good agreement with Faber's $G_0W_0$
values calculated in localized basis sets. However, larger deviations
are clearly observed in some of the lone pair valence states. Their
$G_0W_0$ VIPs are higher than our values by 0.30, 0.40, 0.38, and 0.37
eV for HOMO-1 (the first lone pair state) of cytosine, HOMO-1 (the
first lone pair state) of thymine, and HOMO-1 and HOMO-3 (the first
and second lone pair states) of uracil, respectively. We plot in
Fig.~\ref{FIG:CONVERGENCE2} the convergence behavior of VIPs with
respect to the dimension of the polarizability basis for these lone
pair states to check whether convergence issues are present. But it is
apparent that VIPs from our $G_0W_0$ calculations are fully
converged. Another significant difference is found in the
valence-bound VEAs for all five LUMO levels. Moreover, Faber's
$G_0W_0$ VB-VEAs are lower than the present results by 0.61, 0.39,
0.43, 0.38, and 0.34 eV for G, A, C, T, and U, respectively. It is
interesting to notice that similar trends of increased VIPs and
decreased VEAs are observed in the previous convergence benchmark of
Fig.~\ref{FIG:CONVERGENCE}, when a small optimal polarizability basis
was employed. However, since we do not find significant difference in
the $G_0W_0$ VIPs for other QP states, the source of the above
deviations is not clear. Furthermore, as listed in Table~\ref{TABLE1},
the work by Faber {\em et al.}  demonstrated the importance of
self-consistency of QP energies in $GW$ calculations with QP
wavefunctions unchanged. This self-consistent $GW$ method increases
the $G_0W_0$ VIPs of the HOMO levels by 0.32, 0.32, 0.52, 0.41, and
0.44 eV and decreases the $G_0W_0$ VEAs of the LUMO levels by 0.54,
0.50, 0.46, 0.53, and 0.53 eV for G, A, C, T, and U,
respectively. Results from advanced quantum chemistry methods are also
listed in Table~\ref{TABLE1}, including complete active space with
second-order perturbation theory (CASPT2)~\cite{Roca-SanjuanRMS06,
  Roca-SanjuanMSR08}, coupled-cluster with singles, doubles, and
perturbative triple excitations [CCSD(T)]~\cite{Roca-SanjuanRMS06,
  Roca-SanjuanMSR08}, and equation of motion ionization potential
coupled-cluster (EOM-IP-CCSD)~\cite{BravayaKDLAK10}. VIPs from CASPT2,
CCSD(T), and EOM-IP-CCSD for the HOMO levels are very similar, and
close to the experimental mean values within 0.07, 0.07, and 0.05 eV,
respectively. VEAs from CASPT2 and CCSD(T) for the LUMO levels are
also close to each other; however, they are less close to the mean
experimental values (within 0.30 and 0.33 eV, respectively).  Among
all the theoretical approaches, self-consistent $GW$ and quantum
chemistry methods provide the VIPs and VEAs with the smaller errors
with respect to the experimental data.

\begin{figure}[thpb]\centering
  \includegraphics[width=0.95\columnwidth]{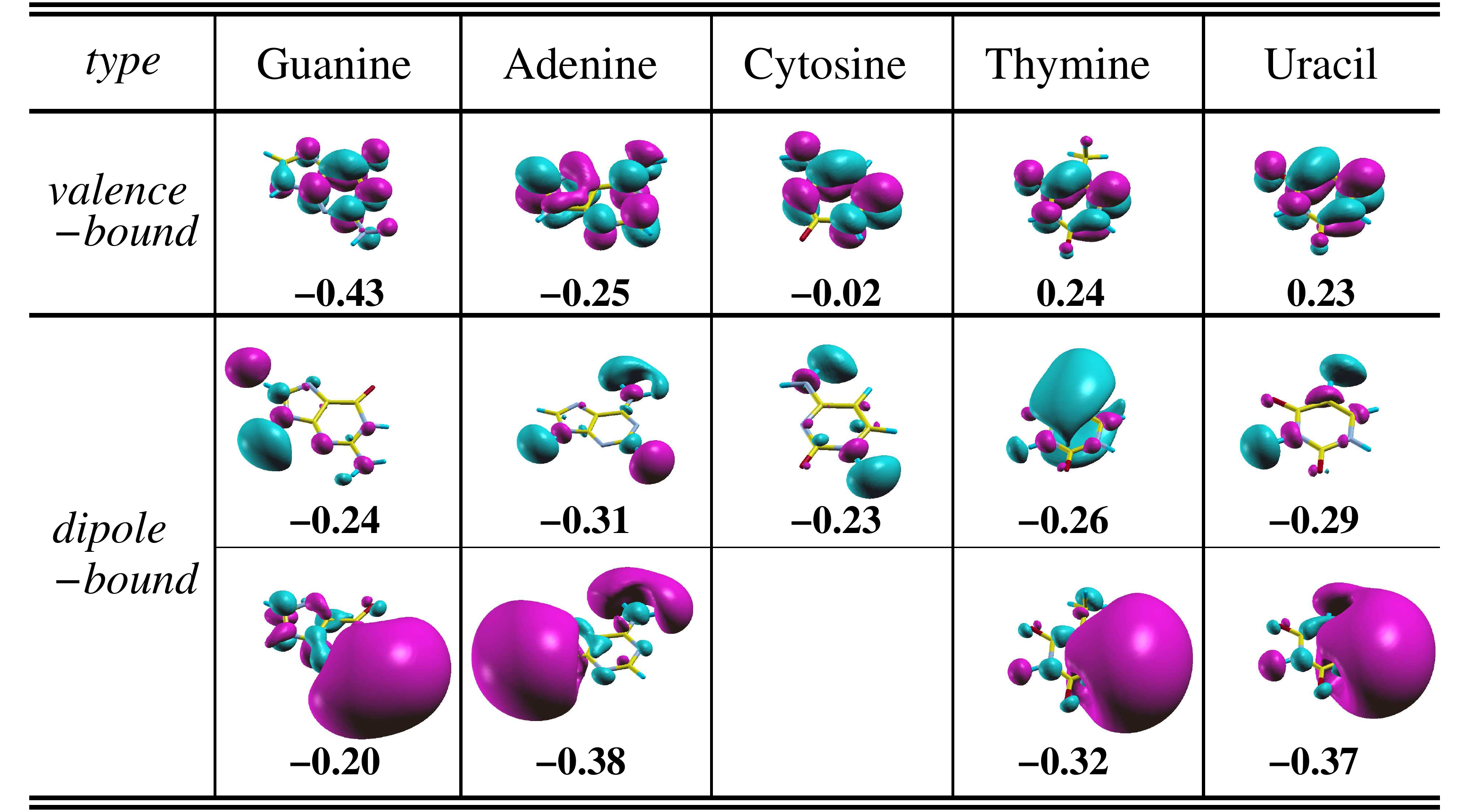}
  \caption{(color online). Valence-bound and dipole-bound VEAs and
    their corresponding QP states, calculated at the DFT-PBE level, in
    the five DNA and RNA bases. Values listed below are VEAs in the
    unit of eV.}
  \label{FIG:EA_ORBITALS}
\end{figure}

Beside the VB-VEAs, there also exist dipole-bound (DB) VEAs, which
correspond to having the additional electron weakly bound to the DNA
and RNA bases by local electrostatic
dipoles~\cite{Roca-SanjuanMSR08}. Both types of QP states are shown in
Fig.~\ref{FIG:EA_ORBITALS}.  It is clear that all five VB states are
localized $\pi^*$ states, while DB states present large lobes, highly
extended outside the molecules. These lobes are mainly located in the
vicinity of the N-H bond, and with a non-negligible dipole moment
along their bond axis.  The energy difference between the VB-VEAs and
their nearest DB-VEAs, $\Delta_{\rm VEA} \equiv$ VEA(VB) $-$ VEA(DB),
are --0.23, 0.06, 0.21, 0.48, and 0.52 eV for G, A, C, T, and U,
respectively.  This suggests that at the $G_0W_0$ level VB states in
the latter four bases are energetically more stable than the DB ones.

\begin{figure}[thpb]\centering
  \includegraphics[width=0.95\columnwidth]{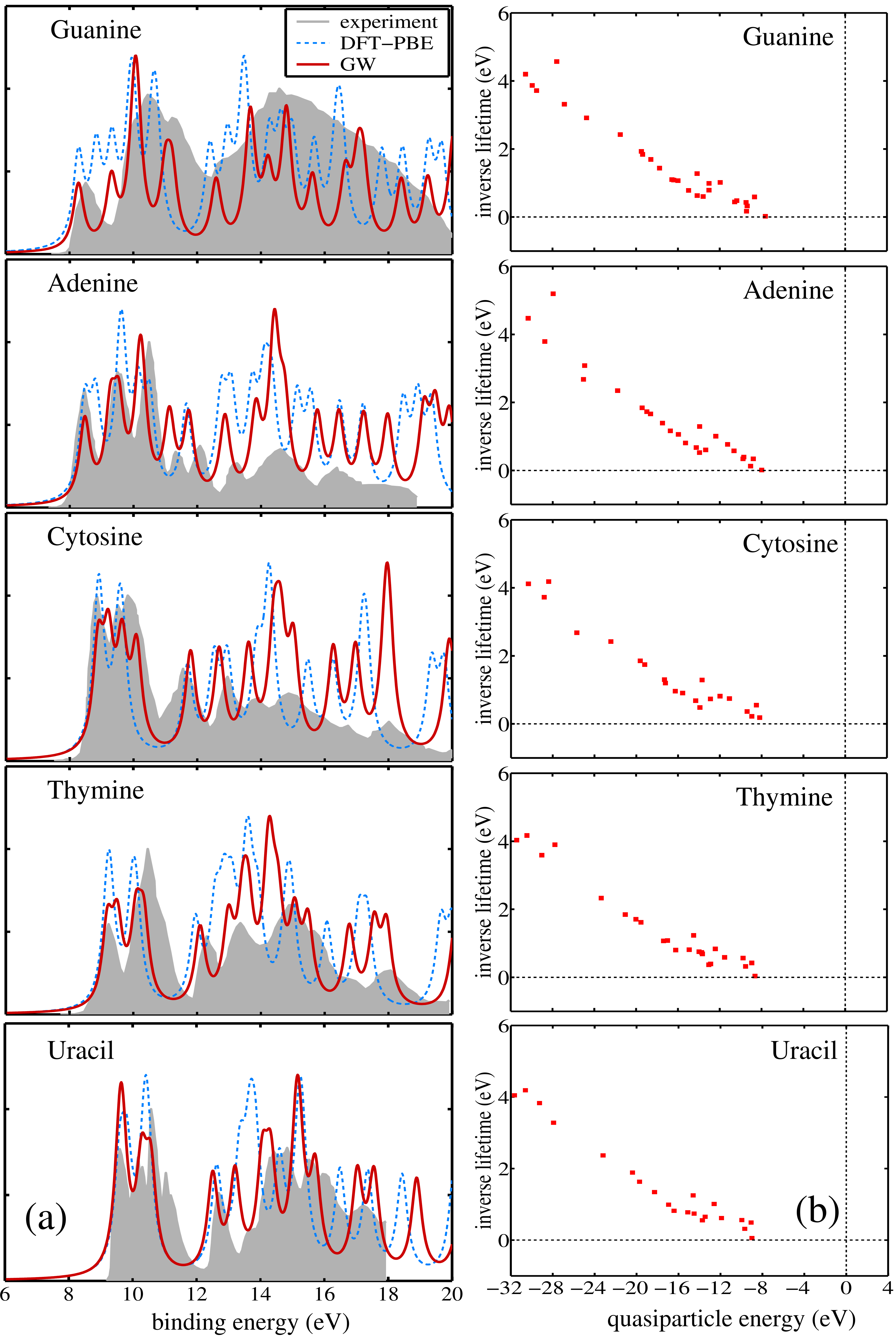}
  \caption{(color online). (a) Experimental valence photoemission
    spectrum (shaded gray area), DFT-PBE DOS (blue dashed lines), and
    $G_0W_0$ DOS (red solid lines). Both DOS curves are shifted to
    match the first VIP of experimental data. Experimental PES spectra
    of G, A, C, T, and U are extracted from
    Refs.~\onlinecite{LinYPALLL80a}, \onlinecite{TrofimovSKPHK06},
    \onlinecite{TrofimovSKPHK06}, \onlinecite{TrofimovSKPHK06}, and
    \onlinecite{OdonnellPSL80}, respectively. The theoretical DOS have
    been obtained through a Lorentzian broadening defined by a width
    of $0.4$ eV. (b) $G_0W_0$ QP energies and inverse lifetime for
    valence states (unit: eV).}
  \label{FIG:PES}
\end{figure}

Experimental valence photoemission spectra extends into deep valence
states~\cite{LinYPALLL80a, OdonnellPSL80, TrofimovSKPHK06}, allowing
us to further evaluate our $G_0W_0$ results at a broader energy
range. The DFT-PBE and $G_0W_0$ densities of states (DOS) for all five
bases, neglecting any oscillator strength effect, are compared to
valence photoemission spectra in Fig.~\ref{FIG:PES}(a). For better
comparison, both curves are shifted to match the first
experimental VIP. It is clearly shown that for all five bases the
$G_0W_0$ DOS agrees much better with the experiment than the DFT-PBE
DOS, thanks to the correct relative position of the various
peaks. Moreover, the $G_0W_0$ self-energy not only leads to large
corrections to DFT eigenvalues, but also provides an estimation of QP
intrinsic lifetimes due to inelastic electron-electron scattering, as
reflected in the imaginary part of QP energies, with $1/\tau_n = 2
|{\rm Im}(\varepsilon_{n}^{\rm QP})|$.  The calculated QP inverse
lifetimes at the $G_0W_0$ level are plotted in Fig.~\ref{FIG:PES}(b)
against the corresponding QP valence energies.  Although $G_0W_0$
permits only a rough estimate of QP lifetimes (the exact ones are
expected to be zero in the range $[2{\rm Re}(\varepsilon_{\rm
  HOMO}^{\rm QP}), \;{\rm Re}(\varepsilon_{\rm HOMO}^{\rm QP})]$), we
note that the QP inverse lifetimes decrease almost linearly with
respect to QP energies for the deep valence states in all five cases.
However, it is still unknown to what extent the $G_0W_0$ estimation of
inverse lifetime would be modified by fully self-consistent $GW$
calculations.

\section{ROLE OF EXCHANGE AND CORRELATION IN GROUND-STATE DFT AND $GW$  CALCULATIONS}
\label{SEC:ROLE}

\begin{figure}[thpb]\centering
  \includegraphics[width=0.99\columnwidth]{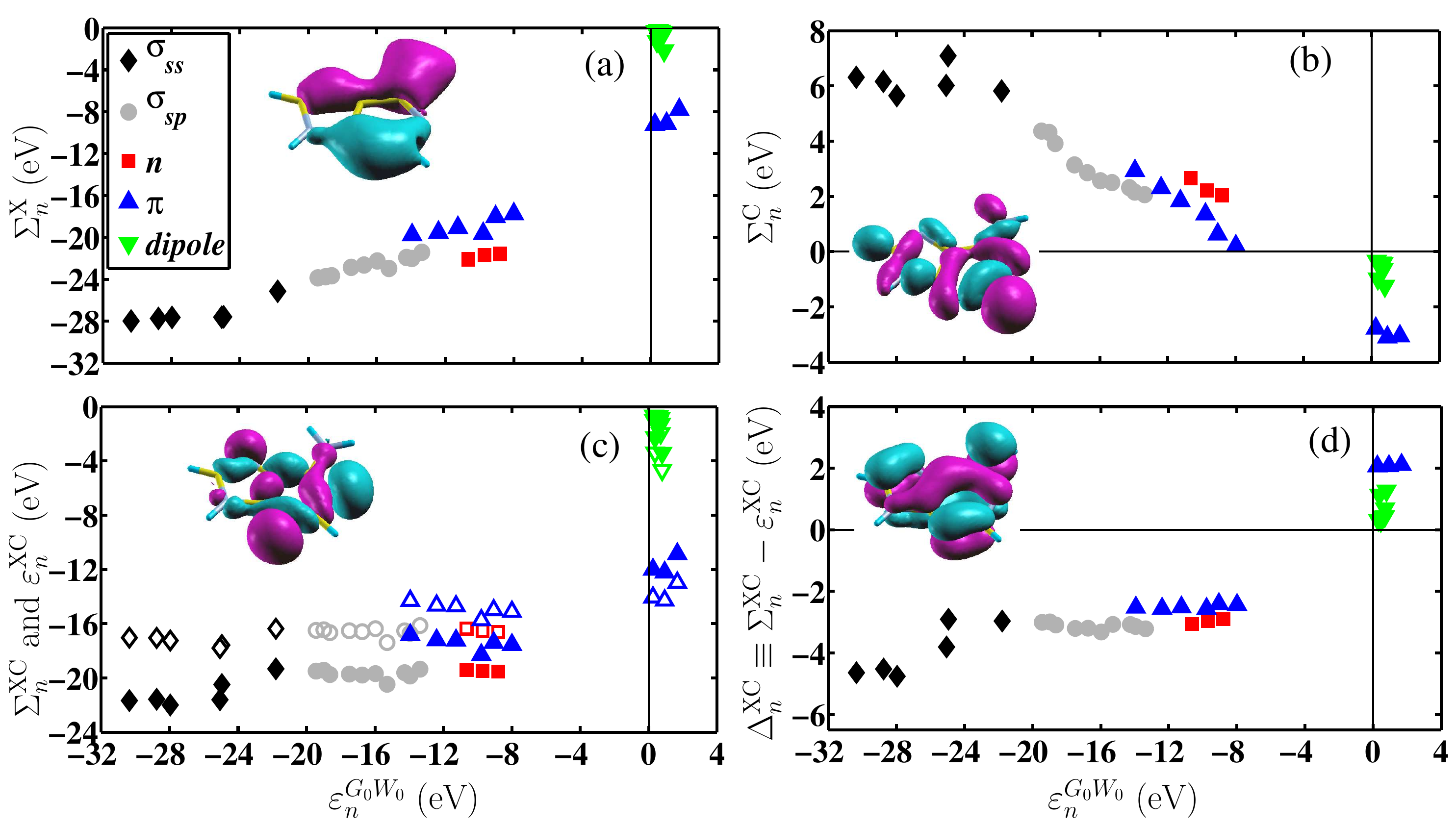}
  \caption{(color online). The role of exchange and correlation in the
    $G_0W_0$ self-energy corrections to Kohn-Sham eigenvalues of 25
    valence states and 10 conduction states in adenine. (a) $G_0W_0$
    exchange energy $\Sigma_n^{\rm X}$, (b) $G_0W_0$ correlation
    energy $\Sigma_n^{\rm C}$, (c) the sum of $G_0W_0$ exchange and
    correlation energy $\Sigma_n^{\rm XC}$ (filled symbols) and DFT XC
    energy $\varepsilon_n^{\rm XC}$ (unfilled symbols), and (d) the
    difference between $G_0W_0$ and DFT exchange-correlation energy,
    $\Delta_n^{\rm XC} \equiv \Sigma_n^{\rm XC} - \varepsilon_n^{\rm
      XC}$. Four types of molecular orbitals are illustrated in
    (a)--(d), corresponding to $\sigma_{ss}$, $\sigma_{sp}$, $n$, and
    $\pi$ characters.}
  \label{FIG:ENERGY_DECOMPOSITION}
\end{figure}

To understand the role of exchange and correlation in the
self-energy corrections to the DFT-PBE results, we first express each
Kohn-Sham eigenvalue $\varepsilon_n^{\rm KS}$ of eigenstate $\psi_n$
for the $n$-th state as the sum of a single-particle energy
$\varepsilon_n^{\rm S}$ and an exchange-correlation energy
$\varepsilon_n^{\rm XC}$: $\varepsilon_n^{\rm KS} = \varepsilon_n^{\rm
  S} + \varepsilon_n^{\rm XC}$, where $\varepsilon_n^{\rm S}$ contains
the energy contributions from the kinetic energy operator, the
external ionic potential, and the Hartree term.  Furthermore, the
$G_0W_0$ QP energy can be written in terms of the exchange self-energy
$\Sigma_n^{\rm X}$ and of the correlation self-energy $\Sigma_n^{\rm
  C}$: $\varepsilon_n^{\rm G_0W_0} = \varepsilon_n^{\rm S} +
\Sigma_n^{\rm X} + \Sigma_n^{\rm C}$.  Exchange and correlation
effects can then be systematically investigated by analyzing
$\Sigma_n^{\rm X}$, $\Sigma_n^{\rm C}$, $\Sigma_n^{\rm XC}$,
$\varepsilon_n^{\rm XC}$, and $\Delta_n^{\rm XC}$, with $\Sigma_n^{\rm XC} \equiv
\Sigma_n^{\rm X}+\Sigma_n^{\rm C}$ and $\Delta_n^{\rm XC}\equiv \Sigma_n^{\rm
  XC} - \varepsilon_n^{\rm XC}$. We consider the adenine molecule
and plot the above quantities with respect to the $G_0W_0$ QP energy
$\varepsilon_n^{G_0W_0}$. As shown in
Figs.~\ref{FIG:ENERGY_DECOMPOSITION}(a) and (b), the $G_0W_0$ exchange
energy $\Sigma_n^{\rm X}$ increases from $-28.0$ to $-17.8$ eV for the
25 valence states and from $-9.2$ to $-0.2$ eV for the 10 conduction
states, while the $G_0W_0$ correlation energy $\Sigma_n^{\rm C}$
decreases from 7.1 down to 0.2 eV for the valence states and from
$-0.3$ to $-3.1$ eV for the conduction states. This clearly shows that
$\Sigma_n^{\rm X}$ is always negative, stabilizing both electron and
hole excitations; however, $\Sigma_n^{\rm C}$ is positive for valence
states and negative for conduction states, indicating that the effect
of correlation is that of destabilizing hole excitations and of
stabilizing electron excitations. Although $\Sigma_n^{\rm X}$ and
$\Sigma_n^{\rm C}$ have opposite trends for hole excitations, exchange
interactions eventually dominate due to their larger magnitude,
leading to the negative $\Sigma_n^{\rm XC}$ of
Fig.~\ref{FIG:ENERGY_DECOMPOSITION}(c).  Interestingly, the $G_0W_0$
$\Sigma_n^{\rm XC}$ is lower than the DFT-PBE $\varepsilon_n^{\rm XC}$
for the valence states, but higher than $\varepsilon_n^{\rm XC}$ for
the conduction states. Consequently, the difference $\Delta_n^{\rm
  XC}$ between $\Sigma_n^{\rm XC}$ and $\varepsilon_n^{\rm XC}$, shown
in Fig.~\ref{FIG:ENERGY_DECOMPOSITION}(d), is negative for
the valence manifold and positive for the conduction manifold,
resulting in an increased HOMO-LUMO gap.  The same behavior is
observed for the other four bases as well.

As shown in Fig.~\ref{FIG:ENERGY_DECOMPOSITION}, we can recognize five
major orbital types among the valence and conduction orbitals of the
isolated adenine molecule: $\sigma_{ss}$, $\sigma_{sp}$, $n$, $\pi$,
and dipole-bound states. The lowest six states correspond to $\sigma$
orbitals due to $s$-$s$ hybridization, which have larger $G_0W_0$
exchange, correlation, and total self-energy corrections than the
other states. The following ten states at higher energy levels exhibit
$\sigma_{sp}$ character, and their $\Sigma_n^{\rm X}$ and
$\Sigma_n^{\rm C}$ show a linear but opposite dependence with respect
to the $G_0W_0$ QP energy $\varepsilon_n^{G_0W_0}$.  Thus, their sum
$\Sigma_n^{\rm XC}$ is shown to be almost constant, ranging from
$-20.5$ to $-19.4$ eV. Since the same trend is present in
$\varepsilon_n^{\rm XC}$, the final difference $\Delta_n^{\rm XC}$
between $G_0W_0$ and DFT results stays almost constant, between $-3.3$
and $-3.0$ eV.  The next three $n$ and six $\pi$ valence states and
three $\pi^*$ conduction states have a similar behavior, despite
different magnitudes in their self-energy corrections. In particular,
the six $\pi$ valence states are lowered by about $-2.5$ eV, while the
three $\pi^*$ conduction states are lifted by 2.1 eV, leading to an
increase of 4.6 eV for the HOMO-LUMO gap.  The above observations
provide an important evidence that the $G_0W_0$ self-energy
corrections are highly orbital-dependent and on average $\Sigma^{\rm
  X}(\sigma_{ss})<\Sigma^{\rm X}(\sigma_{sp}) < \Sigma^{\rm X}(n) <
\Sigma^{\rm X}(\pi)$, $\Sigma^{\rm C}(\sigma_{ss}) > \Sigma^{\rm
  C}(\sigma_{sp}) > \Sigma^{\rm C}(n) > \Sigma^{\rm C}(\pi)$, and
$\Delta^{\rm XC}(\sigma_{ss}) < \Delta^{\rm XC}(\sigma_{sp}) \approx
\Delta^{\rm XC}(n) < \Delta^{\rm XC}(\pi)$.  Consequently, the
commonly-used ``scissor operator'' to correct bandgaps by rigidly
lowering the valence levels and increasing the conduction levels by
the same amount will never be adequate for describing the entire QP
spectrum.

\section{SUMMARY}
\label{SEC:SUMMARY}

In summary, VIPs, VEAs, and DOS of five DNA and RNA bases obtained from
a fully converged many-body $G_0W_0$ approach are found to be in very
good agreement with experiments and other theoretical works. Two types
of vertical electron affinities are found, corresponding to localized
valence-bound excitations and delocalized dipole-bound
excitations. Our calculations further reveal that QP inverse lifetimes
depend linearly on QP energies for the deep valence states. They,
however, come from the zero-th order $G_0W_0$ estimation, and may be
significantly affected in self-consistent $GW$
calculations. Interestingly, the $G_0W_0$ self-energy corrections are
highly orbital dependent, but remain relatively constant for the
states with similar bonding character.  Moreover, $G_0W_0$ VIPs of
lone pair states deviate from the experimental ones more than those
for $\pi$ states. Whether this difference comes from the different
self-interaction errors in Kohn-Sham eigenstates will require further
studies using self-interaction corrected functionals~\cite{PerdewZ81,
  DaboFPLMC10, Korzdorfer11}; work is in progress along this
direction.

\begin{acknowledgments}
  The authors would like to thank Davide Ceresoli and Andrea Ferretti
  for valuable discussions. This work was supported by the Department
  of Energy SciDAC program on Quantum Simulations of Materials and
  Nanostructures (DE-FC02-06ER25794) and Eni S.p.A. under the Eni-MIT
  Alliance Solar Frontiers Program.
\end{acknowledgments}


\begin{thebibliography}{41}
\expandafter\ifx\csname natexlab\endcsname\relax\def\natexlab#1{#1}\fi
\expandafter\ifx\csname bibnamefont\endcsname\relax
  \def\bibnamefont#1{#1}\fi
\expandafter\ifx\csname bibfnamefont\endcsname\relax
  \def\bibfnamefont#1{#1}\fi
\expandafter\ifx\csname citenamefont\endcsname\relax
  \def\citenamefont#1{#1}\fi
\expandafter\ifx\csname url\endcsname\relax
  \def\url#1{\texttt{#1}}\fi
\expandafter\ifx\csname urlprefix\endcsname\relax\def\urlprefix{URL }\fi
\providecommand{\bibinfo}[2]{#2}
\providecommand{\eprint}[2][]{\url{#2}}

\bibitem[{\citenamefont{Colson and Sevilla}(1995)}]{ColsonS95}
\bibinfo{author}{\bibfnamefont{A.~O.} \bibnamefont{Colson}} \bibnamefont{and}
  \bibinfo{author}{\bibfnamefont{M.~D.} \bibnamefont{Sevilla}},
  \bibinfo{journal}{J. Phys. Chem.} \textbf{\bibinfo{volume}{99}},
  \bibinfo{pages}{3867} (\bibinfo{year}{1995}).

\bibitem[{\citenamefont{Zwolak and Di~Ventra}(2005)}]{ZwolakD05}
\bibinfo{author}{\bibfnamefont{M.}~\bibnamefont{Zwolak}} \bibnamefont{and}
  \bibinfo{author}{\bibfnamefont{M.}~\bibnamefont{Di~Ventra}},
  \bibinfo{journal}{Nano Lett.} \textbf{\bibinfo{volume}{5}},
  \bibinfo{pages}{421} (\bibinfo{year}{2005}).

\bibitem[{\citenamefont{Porath et~al.}(2000)\citenamefont{Porath, Bezryadin,
  de~Vries, and Dekker}}]{PorathBdD00}
\bibinfo{author}{\bibfnamefont{D.}~\bibnamefont{Porath}},
  \bibinfo{author}{\bibfnamefont{A.}~\bibnamefont{Bezryadin}},
  \bibinfo{author}{\bibfnamefont{S.}~\bibnamefont{de~Vries}}, \bibnamefont{and}
  \bibinfo{author}{\bibfnamefont{C.}~\bibnamefont{Dekker}},
  \bibinfo{journal}{Nature (London)} \textbf{\bibinfo{volume}{403}},
  \bibinfo{pages}{635} (\bibinfo{year}{2000}).

\bibitem[{\citenamefont{Kawai et~al.}(2009)\citenamefont{Kawai, Kodera,
  Osakada, and Majima}}]{KawaiKOM09}
\bibinfo{author}{\bibfnamefont{K.}~\bibnamefont{Kawai}},
  \bibinfo{author}{\bibfnamefont{H.}~\bibnamefont{Kodera}},
  \bibinfo{author}{\bibfnamefont{Y.}~\bibnamefont{Osakada}}, \bibnamefont{and}
  \bibinfo{author}{\bibfnamefont{T.}~\bibnamefont{Majima}},
  \bibinfo{journal}{Nat. Chem.} \textbf{\bibinfo{volume}{1}},
  \bibinfo{pages}{156} (\bibinfo{year}{2009}).

\bibitem[{\citenamefont{Hush and Cheung}(1975)}]{HushC75}
\bibinfo{author}{\bibfnamefont{N.~S.} \bibnamefont{Hush}} \bibnamefont{and}
  \bibinfo{author}{\bibfnamefont{A.~S.} \bibnamefont{Cheung}},
  \bibinfo{journal}{Chem. Phys. Lett.} \textbf{\bibinfo{volume}{34}},
  \bibinfo{pages}{11} (\bibinfo{year}{1975}).

\bibitem[{\citenamefont{Dougherty et~al.}(1978)\citenamefont{Dougherty,
  Younathan, Voll, Abdulnur, and McGlynn}}]{DoughertyYVAM78}
\bibinfo{author}{\bibfnamefont{D.}~\bibnamefont{Dougherty}},
  \bibinfo{author}{\bibfnamefont{E.~S.} \bibnamefont{Younathan}},
  \bibinfo{author}{\bibfnamefont{R.}~\bibnamefont{Voll}},
  \bibinfo{author}{\bibfnamefont{S.}~\bibnamefont{Abdulnur}}, \bibnamefont{and}
  \bibinfo{author}{\bibfnamefont{S.~P.} \bibnamefont{McGlynn}},
  \bibinfo{journal}{J. Electron Spectrosc. Relat. Phenom.}
  \textbf{\bibinfo{volume}{13}}, \bibinfo{pages}{379} (\bibinfo{year}{1978}).

\bibitem[{\citenamefont{Choi et~al.}(2005)\citenamefont{Choi, Lee, and
  Kim}}]{ChoiLK05}
\bibinfo{author}{\bibfnamefont{K.~W.} \bibnamefont{Choi}},
  \bibinfo{author}{\bibfnamefont{J.~H.} \bibnamefont{Lee}}, \bibnamefont{and}
  \bibinfo{author}{\bibfnamefont{S.~K.} \bibnamefont{Kim}},
  \bibinfo{journal}{J. Am. Chem. Soc.} \textbf{\bibinfo{volume}{127}},
  \bibinfo{pages}{15674} (\bibinfo{year}{2005}).

\bibitem[{\citenamefont{Trofimov et~al.}(2006)\citenamefont{Trofimov, Schirmer,
  Kobychev, Potts, Holland, and Karlsson}}]{TrofimovSKPHK06}
\bibinfo{author}{\bibfnamefont{A.~B.} \bibnamefont{Trofimov}},
  \bibinfo{author}{\bibfnamefont{J.}~\bibnamefont{Schirmer}},
  \bibinfo{author}{\bibfnamefont{V.~B.} \bibnamefont{Kobychev}},
  \bibinfo{author}{\bibfnamefont{A.~W.} \bibnamefont{Potts}},
  \bibinfo{author}{\bibfnamefont{D.~M.~P.} \bibnamefont{Holland}},
  \bibnamefont{and} \bibinfo{author}{\bibfnamefont{L.}~\bibnamefont{Karlsson}},
  \bibinfo{journal}{J. Phys. B-At. Mol. Opt. Phys.}
  \textbf{\bibinfo{volume}{39}}, \bibinfo{pages}{305} (\bibinfo{year}{2006}).

\bibitem[{\citenamefont{Schwell et~al.}(2008)\citenamefont{Schwell, Jochims,
  Baumgartel, and Leach}}]{SchwellJBL08}
\bibinfo{author}{\bibfnamefont{M.}~\bibnamefont{Schwell}},
  \bibinfo{author}{\bibfnamefont{H.~W.} \bibnamefont{Jochims}},
  \bibinfo{author}{\bibfnamefont{H.}~\bibnamefont{Baumgartel}},
  \bibnamefont{and} \bibinfo{author}{\bibfnamefont{S.}~\bibnamefont{Leach}},
  \bibinfo{journal}{Chem. Phys.} \textbf{\bibinfo{volume}{353}},
  \bibinfo{pages}{145} (\bibinfo{year}{2008}).

\bibitem[{\citenamefont{Zaytseva et~al.}(2009)\citenamefont{Zaytseva, Trofimov,
  Schirmer, Plekan, Feyer, Richter, Coreno, and Prince}}]{ZaytsevaTSPFRCP09}
\bibinfo{author}{\bibfnamefont{I.~L.} \bibnamefont{Zaytseva}},
  \bibinfo{author}{\bibfnamefont{A.~B.} \bibnamefont{Trofimov}},
  \bibinfo{author}{\bibfnamefont{J.}~\bibnamefont{Schirmer}},
  \bibinfo{author}{\bibfnamefont{O.}~\bibnamefont{Plekan}},
  \bibinfo{author}{\bibfnamefont{V.}~\bibnamefont{Feyer}},
  \bibinfo{author}{\bibfnamefont{R.}~\bibnamefont{Richter}},
  \bibinfo{author}{\bibfnamefont{M.}~\bibnamefont{Coreno}}, \bibnamefont{and}
  \bibinfo{author}{\bibfnamefont{K.~C.} \bibnamefont{Prince}},
  \bibinfo{journal}{J. Phys. Chem. A} \textbf{\bibinfo{volume}{113}},
  \bibinfo{pages}{15142} (\bibinfo{year}{2009}).

\bibitem[{\citenamefont{Kostko et~al.}(2010)\citenamefont{Kostko, Bravaya,
  Krylov, and Ahmed}}]{KostkoBKA10}
\bibinfo{author}{\bibfnamefont{O.}~\bibnamefont{Kostko}},
  \bibinfo{author}{\bibfnamefont{K.}~\bibnamefont{Bravaya}},
  \bibinfo{author}{\bibfnamefont{A.}~\bibnamefont{Krylov}}, \bibnamefont{and}
  \bibinfo{author}{\bibfnamefont{M.}~\bibnamefont{Ahmed}},
  \bibinfo{journal}{Phys. Chem. Chem. Phys.} \textbf{\bibinfo{volume}{12}},
  \bibinfo{pages}{2860} (\bibinfo{year}{2010}).

\bibitem[{\citenamefont{Russo et~al.}(2000)\citenamefont{Russo, Toscano, and
  Grand}}]{RussoTG00}
\bibinfo{author}{\bibfnamefont{N.}~\bibnamefont{Russo}},
  \bibinfo{author}{\bibfnamefont{M.}~\bibnamefont{Toscano}}, \bibnamefont{and}
  \bibinfo{author}{\bibfnamefont{A.}~\bibnamefont{Grand}}, \bibinfo{journal}{J.
  Comput. Chem.} \textbf{\bibinfo{volume}{21}}, \bibinfo{pages}{1243}
  (\bibinfo{year}{2000}).

\bibitem[{\citenamefont{Roca-Sanjuan et~al.}(2006)\citenamefont{Roca-Sanjuan,
  Rubio, Merchan, and Serrano-Andres}}]{Roca-SanjuanRMS06}
\bibinfo{author}{\bibfnamefont{D.}~\bibnamefont{Roca-Sanjuan}},
  \bibinfo{author}{\bibfnamefont{M.}~\bibnamefont{Rubio}},
  \bibinfo{author}{\bibfnamefont{M.}~\bibnamefont{Merchan}}, \bibnamefont{and}
  \bibinfo{author}{\bibfnamefont{L.}~\bibnamefont{Serrano-Andres}},
  \bibinfo{journal}{J. Chem. Phys.} \textbf{\bibinfo{volume}{125}},
  \bibinfo{pages}{084302} (\bibinfo{year}{2006}).

\bibitem[{\citenamefont{Roca-Sanjuan et~al.}(2008)\citenamefont{Roca-Sanjuan,
  Merchan, Serrano-Andres, and Rubio}}]{Roca-SanjuanMSR08}
\bibinfo{author}{\bibfnamefont{D.}~\bibnamefont{Roca-Sanjuan}},
  \bibinfo{author}{\bibfnamefont{M.}~\bibnamefont{Merchan}},
  \bibinfo{author}{\bibfnamefont{L.}~\bibnamefont{Serrano-Andres}},
  \bibnamefont{and} \bibinfo{author}{\bibfnamefont{M.}~\bibnamefont{Rubio}},
  \bibinfo{journal}{J. Chem. Phys.} \textbf{\bibinfo{volume}{129}},
  \bibinfo{pages}{095104} (\bibinfo{year}{2008}).

\bibitem[{\citenamefont{Bravaya et~al.}(2010)\citenamefont{Bravaya, Kostko,
  Dolgikh, Landau, Ahmed, and Krylov}}]{BravayaKDLAK10}
\bibinfo{author}{\bibfnamefont{K.~B.} \bibnamefont{Bravaya}},
  \bibinfo{author}{\bibfnamefont{O.}~\bibnamefont{Kostko}},
  \bibinfo{author}{\bibfnamefont{S.}~\bibnamefont{Dolgikh}},
  \bibinfo{author}{\bibfnamefont{A.}~\bibnamefont{Landau}},
  \bibinfo{author}{\bibfnamefont{M.}~\bibnamefont{Ahmed}}, \bibnamefont{and}
  \bibinfo{author}{\bibfnamefont{A.~I.} \bibnamefont{Krylov}},
  \bibinfo{journal}{J. Phys. Chem. A} \textbf{\bibinfo{volume}{114}},
  \bibinfo{pages}{12305} (\bibinfo{year}{2010}).

\bibitem[{\citenamefont{Hedin}(1965)}]{Hedin65}
\bibinfo{author}{\bibfnamefont{L.}~\bibnamefont{Hedin}},
  \bibinfo{journal}{Phys. Rev.} \textbf{\bibinfo{volume}{139}},
  \bibinfo{pages}{A796} (\bibinfo{year}{1965}).

\bibitem[{\citenamefont{Hedin and Lundqvist}(1969)}]{HedinL69}
\bibinfo{author}{\bibfnamefont{L.}~\bibnamefont{Hedin}} \bibnamefont{and}
  \bibinfo{author}{\bibfnamefont{S.}~\bibnamefont{Lundqvist}}, in
  \emph{\bibinfo{booktitle}{Solid State Physics, Advances in Research and
  Application}}, edited by
  \bibinfo{editor}{\bibfnamefont{F.}~\bibnamefont{Seitz}},
  \bibinfo{editor}{\bibfnamefont{D.}~\bibnamefont{Turnbull}}, \bibnamefont{and}
  \bibinfo{editor}{\bibfnamefont{H.}~\bibnamefont{Ehrenreich}}
  (\bibinfo{publisher}{Academic Press}, \bibinfo{address}{New York},
  \bibinfo{year}{1969}), vol.~\bibinfo{volume}{23}, pp.
  \bibinfo{pages}{1--181}.

\bibitem[{\citenamefont{Hybertsen and Louie}(1986)}]{HybertsenL86}
\bibinfo{author}{\bibfnamefont{M.~S.} \bibnamefont{Hybertsen}}
  \bibnamefont{and} \bibinfo{author}{\bibfnamefont{S.~G.} \bibnamefont{Louie}},
  \bibinfo{journal}{Phys. Rev. B} \textbf{\bibinfo{volume}{34}},
  \bibinfo{pages}{5390} (\bibinfo{year}{1986}).

\bibitem[{\citenamefont{Rojas et~al.}(1995)\citenamefont{Rojas, Godby, and
  Needs}}]{RojasGN95}
\bibinfo{author}{\bibfnamefont{H.~N.} \bibnamefont{Rojas}},
  \bibinfo{author}{\bibfnamefont{R.~W.} \bibnamefont{Godby}}, \bibnamefont{and}
  \bibinfo{author}{\bibfnamefont{R.~J.} \bibnamefont{Needs}},
  \bibinfo{journal}{Phys. Rev. Lett.} \textbf{\bibinfo{volume}{74}},
  \bibinfo{pages}{1827} (\bibinfo{year}{1995}).

\bibitem[{\citenamefont{Rieger et~al.}(1999)\citenamefont{Rieger, Steinbeck,
  White, Rojas, and Godby}}]{RiegerSWRG99}
\bibinfo{author}{\bibfnamefont{M.~M.} \bibnamefont{Rieger}},
  \bibinfo{author}{\bibfnamefont{L.}~\bibnamefont{Steinbeck}},
  \bibinfo{author}{\bibfnamefont{I.~D.} \bibnamefont{White}},
  \bibinfo{author}{\bibfnamefont{H.~N.} \bibnamefont{Rojas}}, \bibnamefont{and}
  \bibinfo{author}{\bibfnamefont{R.~W.} \bibnamefont{Godby}},
  \bibinfo{journal}{Comput. Phys. Commun.} \textbf{\bibinfo{volume}{117}},
  \bibinfo{pages}{211} (\bibinfo{year}{1999}).

\bibitem[{\citenamefont{Campillo et~al.}(1999)\citenamefont{Campillo, Pitarke,
  Rubio, Zarate, and Echenique}}]{CampilloPRZE99}
\bibinfo{author}{\bibfnamefont{I.}~\bibnamefont{Campillo}},
  \bibinfo{author}{\bibfnamefont{J.~M.} \bibnamefont{Pitarke}},
  \bibinfo{author}{\bibfnamefont{A.}~\bibnamefont{Rubio}},
  \bibinfo{author}{\bibfnamefont{E.}~\bibnamefont{Zarate}}, \bibnamefont{and}
  \bibinfo{author}{\bibfnamefont{P.~M.} \bibnamefont{Echenique}},
  \bibinfo{journal}{Phys. Rev. Lett.} \textbf{\bibinfo{volume}{83}},
  \bibinfo{pages}{2230} (\bibinfo{year}{1999}).

\bibitem[{\citenamefont{Spataru et~al.}(2001)\citenamefont{Spataru, Cazalilla,
  Rubio, Benedict, Echenique, and Louie}}]{SpataruCRBEL01}
\bibinfo{author}{\bibfnamefont{C.~D.} \bibnamefont{Spataru}},
  \bibinfo{author}{\bibfnamefont{M.~A.} \bibnamefont{Cazalilla}},
  \bibinfo{author}{\bibfnamefont{A.}~\bibnamefont{Rubio}},
  \bibinfo{author}{\bibfnamefont{L.~X.} \bibnamefont{Benedict}},
  \bibinfo{author}{\bibfnamefont{P.~M.} \bibnamefont{Echenique}},
  \bibnamefont{and} \bibinfo{author}{\bibfnamefont{S.~G.} \bibnamefont{Louie}},
  \bibinfo{journal}{Phys. Rev. Lett.} \textbf{\bibinfo{volume}{87}},
  \bibinfo{pages}{246405} (\bibinfo{year}{2001}).

\bibitem[{\citenamefont{Onida et~al.}(2002)\citenamefont{Onida, Reining, and
  Rubio}}]{OnidaRR02}
\bibinfo{author}{\bibfnamefont{G.}~\bibnamefont{Onida}},
  \bibinfo{author}{\bibfnamefont{L.}~\bibnamefont{Reining}}, \bibnamefont{and}
  \bibinfo{author}{\bibfnamefont{A.}~\bibnamefont{Rubio}},
  \bibinfo{journal}{Rev. Mod. Phys.} \textbf{\bibinfo{volume}{74}},
  \bibinfo{pages}{601} (\bibinfo{year}{2002}).

\bibitem[{\citenamefont{Dori et~al.}(2006)\citenamefont{Dori, Menon, Kilian,
  Sokolowski, Kronik, and Umbach}}]{DoriMKSKU06}
\bibinfo{author}{\bibfnamefont{N.}~\bibnamefont{Dori}},
  \bibinfo{author}{\bibfnamefont{M.}~\bibnamefont{Menon}},
  \bibinfo{author}{\bibfnamefont{L.}~\bibnamefont{Kilian}},
  \bibinfo{author}{\bibfnamefont{M.}~\bibnamefont{Sokolowski}},
  \bibinfo{author}{\bibfnamefont{L.}~\bibnamefont{Kronik}}, \bibnamefont{and}
  \bibinfo{author}{\bibfnamefont{E.}~\bibnamefont{Umbach}},
  \bibinfo{journal}{Phys. Rev. B} \textbf{\bibinfo{volume}{73}},
  \bibinfo{pages}{195208} (\bibinfo{year}{2006}).

\bibitem[{\citenamefont{Tiago et~al.}(2008)\citenamefont{Tiago, Kent, Hood, and
  Reboredo}}]{TiagoKHR08}
\bibinfo{author}{\bibfnamefont{M.~L.} \bibnamefont{Tiago}},
  \bibinfo{author}{\bibfnamefont{P.~R.~C.} \bibnamefont{Kent}},
  \bibinfo{author}{\bibfnamefont{R.~Q.} \bibnamefont{Hood}}, \bibnamefont{and}
  \bibinfo{author}{\bibfnamefont{F.~A.} \bibnamefont{Reboredo}},
  \bibinfo{journal}{J. Chem. Phys.} \textbf{\bibinfo{volume}{129}},
  \bibinfo{pages}{084311} (\bibinfo{year}{2008}).

\bibitem[{\citenamefont{Palummo et~al.}(2009)\citenamefont{Palummo, Hogan,
  Sottile, Bagala, and Rubio}}]{PalummoHSBR09}
\bibinfo{author}{\bibfnamefont{M.}~\bibnamefont{Palummo}},
  \bibinfo{author}{\bibfnamefont{C.}~\bibnamefont{Hogan}},
  \bibinfo{author}{\bibfnamefont{F.}~\bibnamefont{Sottile}},
  \bibinfo{author}{\bibfnamefont{P.}~\bibnamefont{Bagala}}, \bibnamefont{and}
  \bibinfo{author}{\bibfnamefont{A.}~\bibnamefont{Rubio}}, \bibinfo{journal}{J.
  Chem. Phys.} \textbf{\bibinfo{volume}{131}}, \bibinfo{pages}{084102}
  (\bibinfo{year}{2009}).

\bibitem[{\citenamefont{Umari et~al.}(2009)\citenamefont{Umari, Stenuit, and
  Baroni}}]{UmariSB09}
\bibinfo{author}{\bibfnamefont{P.}~\bibnamefont{Umari}},
  \bibinfo{author}{\bibfnamefont{G.}~\bibnamefont{Stenuit}}, \bibnamefont{and}
  \bibinfo{author}{\bibfnamefont{S.}~\bibnamefont{Baroni}},
  \bibinfo{journal}{Phys. Rev. B} \textbf{\bibinfo{volume}{79}},
  \bibinfo{pages}{201104} (\bibinfo{year}{2009}).

\bibitem[{\citenamefont{Umari et~al.}(2010)\citenamefont{Umari, Stenuit, and
  Baroni}}]{UmariSB10}
\bibinfo{author}{\bibfnamefont{P.}~\bibnamefont{Umari}},
  \bibinfo{author}{\bibfnamefont{G.}~\bibnamefont{Stenuit}}, \bibnamefont{and}
  \bibinfo{author}{\bibfnamefont{S.}~\bibnamefont{Baroni}},
  \bibinfo{journal}{Phys. Rev. B} \textbf{\bibinfo{volume}{81}},
  \bibinfo{pages}{115104} (\bibinfo{year}{2010}).

\bibitem[{\citenamefont{Stenuit et~al.}(2010)\citenamefont{Stenuit,
  Castellarin-Cudia, Plekan, Feyer, Prince, Goldoni, and
  Umari}}]{StenuitCPFPGU10}
\bibinfo{author}{\bibfnamefont{G.}~\bibnamefont{Stenuit}},
  \bibinfo{author}{\bibfnamefont{C.}~\bibnamefont{Castellarin-Cudia}},
  \bibinfo{author}{\bibfnamefont{O.}~\bibnamefont{Plekan}},
  \bibinfo{author}{\bibfnamefont{V.}~\bibnamefont{Feyer}},
  \bibinfo{author}{\bibfnamefont{K.~C.} \bibnamefont{Prince}},
  \bibinfo{author}{\bibfnamefont{A.}~\bibnamefont{Goldoni}}, \bibnamefont{and}
  \bibinfo{author}{\bibfnamefont{P.}~\bibnamefont{Umari}},
  \bibinfo{journal}{Phys. Chem. Chem. Phys.} \textbf{\bibinfo{volume}{12}},
  \bibinfo{pages}{10812} (\bibinfo{year}{2010}).

\bibitem[{\citenamefont{Umari et~al.}(2011)\citenamefont{Umari, Qian, Marzari,
  Stenuit, Giacomazzi, and Baroni}}]{UmariQMSGB11}
\bibinfo{author}{\bibfnamefont{P.}~\bibnamefont{Umari}},
  \bibinfo{author}{\bibfnamefont{X.}~\bibnamefont{Qian}},
  \bibinfo{author}{\bibfnamefont{N.}~\bibnamefont{Marzari}},
  \bibinfo{author}{\bibfnamefont{G.}~\bibnamefont{Stenuit}},
  \bibinfo{author}{\bibfnamefont{L.}~\bibnamefont{Giacomazzi}},
  \bibnamefont{and} \bibinfo{author}{\bibfnamefont{S.}~\bibnamefont{Baroni}},
  \bibinfo{journal}{Phys. Status Solidi (b)} \textbf{\bibinfo{volume}{248}},
  \bibinfo{pages}{527} (\bibinfo{year}{2011}).

\bibitem[{\citenamefont{Rostgaard et~al.}(2010)\citenamefont{Rostgaard,
  Jacobsen, and Thygesen}}]{RostgaardJT10}
\bibinfo{author}{\bibfnamefont{C.}~\bibnamefont{Rostgaard}},
  \bibinfo{author}{\bibfnamefont{K.~W.} \bibnamefont{Jacobsen}},
  \bibnamefont{and} \bibinfo{author}{\bibfnamefont{K.~S.}
  \bibnamefont{Thygesen}}, \bibinfo{journal}{Phys. Rev. B}
  \textbf{\bibinfo{volume}{81}}, \bibinfo{pages}{085103}
  (\bibinfo{year}{2010}).

\bibitem[{\citenamefont{Blase et~al.}(2011)\citenamefont{Blase, Attaccalite,
  and Olevano}}]{BlaseAO11}
\bibinfo{author}{\bibfnamefont{X.}~\bibnamefont{Blase}},
  \bibinfo{author}{\bibfnamefont{C.}~\bibnamefont{Attaccalite}},
  \bibnamefont{and} \bibinfo{author}{\bibfnamefont{V.}~\bibnamefont{Olevano}},
  \bibinfo{journal}{Phys. Rev. B} \textbf{\bibinfo{volume}{83}},
  \bibinfo{pages}{115103} (\bibinfo{year}{2011}).

\bibitem[{\citenamefont{Faber et~al.}(2011)\citenamefont{Faber, Attaccalite,
  Olevano, Runge, and Blase}}]{FaberAORB11}
\bibinfo{author}{\bibfnamefont{C.}~\bibnamefont{Faber}},
  \bibinfo{author}{\bibfnamefont{C.}~\bibnamefont{Attaccalite}},
  \bibinfo{author}{\bibfnamefont{V.}~\bibnamefont{Olevano}},
  \bibinfo{author}{\bibfnamefont{E.}~\bibnamefont{Runge}}, \bibnamefont{and}
  \bibinfo{author}{\bibfnamefont{X.}~\bibnamefont{Blase}},
  \bibinfo{journal}{Phys. Rev. B} \textbf{\bibinfo{volume}{83}},
  \bibinfo{pages}{115123} (\bibinfo{year}{2011}).

\bibitem[{\citenamefont{Giannozzi et~al.}(2009)\citenamefont{Giannozzi, Baroni,
  Bonini, Calandra, Car, Cavazzoni, Ceresoli, Chiarotti, Cococcioni, Dabo
  et~al.}}]{ESPRESSO10}
\bibinfo{author}{\bibfnamefont{P.}~\bibnamefont{Giannozzi}},
  \bibinfo{author}{\bibfnamefont{S.}~\bibnamefont{Baroni}},
  \bibinfo{author}{\bibfnamefont{N.}~\bibnamefont{Bonini}},
  \bibinfo{author}{\bibfnamefont{M.}~\bibnamefont{Calandra}},
  \bibinfo{author}{\bibfnamefont{R.}~\bibnamefont{Car}},
  \bibinfo{author}{\bibfnamefont{C.}~\bibnamefont{Cavazzoni}},
  \bibinfo{author}{\bibfnamefont{D.}~\bibnamefont{Ceresoli}},
  \bibinfo{author}{\bibfnamefont{G.~L.} \bibnamefont{Chiarotti}},
  \bibinfo{author}{\bibfnamefont{M.}~\bibnamefont{Cococcioni}},
  \bibinfo{author}{\bibfnamefont{I.}~\bibnamefont{Dabo}}, \bibnamefont{et~al.},
  \bibinfo{journal}{J. Phys.-Condens. Matter} \textbf{\bibinfo{volume}{21}},
  \bibinfo{pages}{395502} (\bibinfo{year}{2009}),
  \urlprefix\url{http://www.quantum-espresso.org}.

\bibitem[{\citenamefont{Godby et~al.}(1986)\citenamefont{Godby, Schluter, and
  Sham}}]{GodbySS86}
\bibinfo{author}{\bibfnamefont{R.~W.} \bibnamefont{Godby}},
  \bibinfo{author}{\bibfnamefont{M.}~\bibnamefont{Schluter}}, \bibnamefont{and}
  \bibinfo{author}{\bibfnamefont{L.~J.} \bibnamefont{Sham}},
  \bibinfo{journal}{Phys. Rev. Lett.} \textbf{\bibinfo{volume}{56}},
  \bibinfo{pages}{2415} (\bibinfo{year}{1986}).

\bibitem[{\citenamefont{Gruning et~al.}(2006)\citenamefont{Gruning, Marini, and
  Rubio}}]{GruningMR06}
\bibinfo{author}{\bibfnamefont{M.}~\bibnamefont{Gruning}},
  \bibinfo{author}{\bibfnamefont{A.}~\bibnamefont{Marini}}, \bibnamefont{and}
  \bibinfo{author}{\bibfnamefont{A.}~\bibnamefont{Rubio}},
  \bibinfo{journal}{Phys. Rev. B} \textbf{\bibinfo{volume}{74}},
  \bibinfo{pages}{161103} (\bibinfo{year}{2006}).

\bibitem[{\citenamefont{Lin et~al.}(1980)\citenamefont{Lin, Yu, Peng, Akiyama,
  Li, Lee, and Lebreton}}]{LinYPALLL80a}
\bibinfo{author}{\bibfnamefont{J.}~\bibnamefont{Lin}},
  \bibinfo{author}{\bibfnamefont{C.}~\bibnamefont{Yu}},
  \bibinfo{author}{\bibfnamefont{S.}~\bibnamefont{Peng}},
  \bibinfo{author}{\bibfnamefont{I.}~\bibnamefont{Akiyama}},
  \bibinfo{author}{\bibfnamefont{K.}~\bibnamefont{Li}},
  \bibinfo{author}{\bibfnamefont{L.~K.} \bibnamefont{Lee}}, \bibnamefont{and}
  \bibinfo{author}{\bibfnamefont{P.~R.} \bibnamefont{LeBreton}},
  \bibinfo{journal}{J. Phys. Chem.} \textbf{\bibinfo{volume}{84}},
  \bibinfo{pages}{1006} (\bibinfo{year}{1980}).

\bibitem[{\citenamefont{Odonnell et~al.}(1980)\citenamefont{Odonnell, Petke,
  Shipman, and Lebreton}}]{OdonnellPSL80}
\bibinfo{author}{\bibfnamefont{T.~J.} \bibnamefont{O'Donnell}},
  \bibinfo{author}{\bibfnamefont{P.~R.} \bibnamefont{LeBreton}},
  \bibinfo{author}{\bibfnamefont{J.~D.} \bibnamefont{Petke}},
  \bibnamefont{and} \bibinfo{author}{\bibfnamefont{L.~L.}
  \bibnamefont{Shipman}}, \bibinfo{journal}{J. Phys. Chem.}
  \textbf{\bibinfo{volume}{84}}, \bibinfo{pages}{1975} (\bibinfo{year}{1980}).

\bibitem[{\citenamefont{Perdew and Zunger}(1981)}]{PerdewZ81}
\bibinfo{author}{\bibfnamefont{J.~P.} \bibnamefont{Perdew}} \bibnamefont{and}
  \bibinfo{author}{\bibfnamefont{A.}~\bibnamefont{Zunger}},
  \bibinfo{journal}{Phys. Rev. B} \textbf{\bibinfo{volume}{23}},
  \bibinfo{pages}{5048} (\bibinfo{year}{1981}).

\bibitem[{\citenamefont{Dabo et~al.}(2010)\citenamefont{Dabo, Ferretti,
  Poilvert, Li, Marzari, and Cococcioni}}]{DaboFPLMC10}
\bibinfo{author}{\bibfnamefont{I.}~\bibnamefont{Dabo}},
  \bibinfo{author}{\bibfnamefont{A.}~\bibnamefont{Ferretti}},
  \bibinfo{author}{\bibfnamefont{N.}~\bibnamefont{Poilvert}},
  \bibinfo{author}{\bibfnamefont{Y.~L.} \bibnamefont{Li}},
  \bibinfo{author}{\bibfnamefont{N.}~\bibnamefont{Marzari}}, \bibnamefont{and}
  \bibinfo{author}{\bibfnamefont{M.}~\bibnamefont{Cococcioni}},
  \bibinfo{journal}{Phys. Rev. B} \textbf{\bibinfo{volume}{82}},
  \bibinfo{pages}{115121} (\bibinfo{year}{2010}).

\bibitem[{\citenamefont{Korzdorfer}(2011)}]{Korzdorfer11}
\bibinfo{author}{\bibfnamefont{T.}~\bibnamefont{Korzdorfer}},
  \bibinfo{journal}{J. Chem. Phys.} \textbf{\bibinfo{volume}{134}},
  \bibinfo{pages}{094111} (\bibinfo{year}{2011}).

\end{thebibliography}

\end{document}